\documentclass[twocolumn,trackchanges]{aastex61}

\usepackage{graphicx}
\usepackage{epsfig}
\usepackage{times}
\usepackage{natbib}
\usepackage{url}
\usepackage{color}
\usepackage{amssymb,amsmath}
\usepackage{gensymb}

\hypersetup{linkcolor=red,citecolor=cyan,filecolor=cyan,urlcolor=blue}
   
\shorttitle{MHD Simulation of the Bastille Day Event}
\shortauthors{T\"or\"ok et al.}

\begin{document}

\title{SUN-TO-EARTH MHD SIMULATION OF THE 14 JULY 2000 ``BASTILLE DAY'' ERUPTION}

\author[0000-0003-3843-3242]{Tibor T\"or\"ok}
\affiliation{Predictive Science Inc., 9990 Mesa Rim Road, Suite 170, San Diego, CA 92121, USA}
\author[0000-0003-1759-4354]{Cooper Downs}
\affiliation{Predictive Science Inc., 9990 Mesa Rim Road, Suite 170, San Diego, CA 92121, USA}
\author{Jon A. Linker}
\affiliation{Predictive Science Inc., 9990 Mesa Rim Road, Suite 170, San Diego, CA 92121, USA}
\author[0000-0001-9231-045X]{R. Lionello}
\affiliation{Predictive Science Inc., 9990 Mesa Rim Road, Suite 170, San Diego, CA 92121, USA}
\author[0000-0001-7053-4081]{Viacheslav S. Titov}
\affiliation{Predictive Science Inc., 9990 Mesa Rim Road, Suite 170, San Diego, CA 92121, USA}
\author{Zoran Miki\'c}
\affiliation{Predictive Science Inc., 9990 Mesa Rim Road, Suite 170, San Diego, CA 92121, USA}
\author[0000-0002-1859-456X]{Pete Riley}
\affiliation{Predictive Science Inc., 9990 Mesa Rim Road, Suite 170, San Diego, CA 92121, USA}
\author{Ronald M. Caplan}
\affiliation{Predictive Science Inc., 9990 Mesa Rim Road, Suite 170, San Diego, CA 92121, USA}
\author{Janvier Wijaya}
\affiliation{Predictive Science Inc., 9990 Mesa Rim Road, Suite 170, San Diego, CA 92121, USA}
\correspondingauthor{Tibor T\"or\"ok}
\email{tibor@predsci.com}

\begin{abstract}
Solar eruptions are the main driver of space-weather disturbances at the Earth. Extreme events are of particular interest, not only because of the scientific challenges they pose, but also because of their possible societal consequences. Here we present a magnetohydrodynamic (MHD) simulation of the 14 July 2000 ``Bastille Day'' eruption, which produced a very strong geomagnetic storm. After constructing a ``thermodynamic'' MHD model of the corona and solar wind, we insert a magnetically stable flux rope along the polarity inversion line of the eruption's source region and initiate the eruption by boundary flows. More than $10^{33}$ ergs of magnetic energy are released in the eruption within a few minutes, driving a flare, an EUV wave, and a coronal mass ejection (CME) that travels in the outer corona at $\approx$\,1500\,km\,s$^{-1}$, close to the observed speed. We then propagate the CME to Earth, using a heliospheric MHD code. Our simulation thus provides the opportunity to test how well {\em in situ} observations of extreme events are matched if the eruption is initiated from a stable magnetic-equilibrium state. We find that the flux-rope center is very similar in character to the observed magnetic cloud, but arrives $\approx$\,8.5 hours later and $\approx$\,15\degree~too far to the North, with field strengths that are too weak by a factor of $\approx$\,1.6. The front of the flux rope is highly distorted, exhibiting localized magnetic-field concentrations as it passes 1\,AU. We discuss these properties with regard to the development of space-weather predictions based on MHD simulations of solar eruptions.  
\end{abstract}


%%%%%%%%%%%%%%%%%%%%%%%%%%%%%%%%%%%%%%%%%%%%%%%%%%%%%%%
%%%%%%%%%%%%%%%%%%%%%%%%%%%%%%%%%%%%%%%%%%%%%%%%%%%%%%%
%
\section{Introduction}
\label{s:intro}
%
%%%%%%%%%%%%%%%%%%%%%%%%%%%%%%%%%%%%%%%%%%%%%%%%%%%%%%%
%%%%%%%%%%%%%%%%%%%%%%%%%%%%%%%%%%%%%%%%%%%%%%%%%%%%%%%
%
Coronal Mass Ejections (CMEs) are immense eruptions that propel plasma and magnetic flux outward from the Sun.  CMEs (and accompanying solar flares) are the largest impulsive energy release events in the solar system, and are therefore of inherent scientific interest. There are many open scientific questions about such events, such as: how is the energy released so impulsively and how are CMEs initiated? 

The strongest solar eruptions are typically characterized by very fast ($>1000$\,km\,s$^{-1}$) CMEs and X-class solar flares, such as the famous ``Bastille Day'' event considered in this article. Such ``extreme'' eruptions are responsible for the most severe space-weather effects at Earth. Fast CMEs are the primary cause of major geomagnetic storms and are typically associated with solar energetic particle (SEP) events \citep[e.g.,][]{gopalswamy06}, both of which can represent a significant hazard for humans and technological infrastructure \citep[e.g.,][]{baker16}.

Particles accelerated during a flare or CME can reach Earth within half an hour or less \citep[e.g.,][]{schwadron14}, leaving little time for a quantitative prediction of their consequences. On the other hand, even the fastest CMEs require almost a day to arrive at Earth and initiate a geomagnetic storm. This provides, in principle, sufficient time to predict their impact. The geo-effectiveness of CMEs, i.e., of the associated interplanetary CME (ICME) or magnetic cloud (MC), depends primarily on their Earth-side magnetic-field direction (``$B_z$''), their velocity, and on their associated ram pressure upon arrival at the magnetosphere \citep[e.g.,][]{srivastava04,gopalswamy08}. It is therefore highly desirable to predict these parameters before an ICME (and the shock that potentially precedes it) arrives at Earth \citep[e.g.,][]{siscoe06,messerotti09}. A candidate tool for this purpose are magnetohydrodynamic (MHD) numerical simulations.

MHD simulations have been widely employed to model CMEs. Many of them use idealized configurations and are primarily intended to investigate specific aspects of CMEs, such as their initiation mechanisms \citep[e.g.,][]{forbes90,mikic94,amari96,amari00,amari03a,antiochos99a,chen00,fan03,kusano04,kusano12,lynch05,torok05,torok07,aulanier10,karpen12}. For reviews on such simulations and the underlying theoretical concepts see, e.g., \cite{forbes06,aulanier14,green18}. 

Other simulations are specifically designed to model observed events. These models typically derive boundary conditions for the magnetic field from observed magnetograms, and produce the pre-eruptive configuration using boundary flows, nonlinear force-free field (NLFFF) extrapolations, or analytical flux-rope models that are inserted into the source region of the eruption. Some of these simulations just model the initiation and coronal evolution of CMEs \citep[e.g.,][]{roussev07,lugaz07,lugaz09,cohen09,zuccarello12,kliem13,amari14,jiang.c16,fan16}, while others include the propagation of the associated ICME to one astronomical unit (AU) or beyond \citep[e.g.,][]{manchester04b,toth07,shen.f14}. In some cases, the modeling of the eruption is fully neglected, and simplified initial conditions for the CME or ICME (sometimes merely a velocity perturbation) are set up at some distances from the Sun in the corona or in the inner heliosphere \citep[e.g.,][]{shen.f14,zhou.y14,shiota16}.

At the present time, the most advanced simulations of observed eruptions additionally use a sophisticated treatment of the energy transfer in the corona that includes thermal conduction, radiative losses, and empirical (or wave-turbulence driven) coronal heating, which is often referred to as ``thermodynamic MHD'' \citep{lionello01,linker01,lionello09,downs10,downs13,vanderholst10,vanderholst14,sokolov13}. This description is required for modeling the plasma properties in the corona to a degree of realism that allows one to produce synthetic satellite images that can be directly compared to observations \citep{lionello09}. CMEs are then launched in this background environment in the source region of the eruption \citep[e.g.,][]{lugaz11,downs11,downs12,manchester12,jin.m13,jin.m16}, and sometimes coupled to a simpler heliospheric MHD model to propagate the associated ICME to 1\,AU or beyond \citep[e.g.,][]{lionello13a,manchester14,merkin16,jin.m17b}. 

Thermodynamic MHD simulations are complex and computationally expensive, and therefore have not yet been used for operational space-weather predictions, even when coupled to computationally more efficient heliospheric simulations. At present, the only three-dimensional MHD model that is used for operational forecasts (at the NOAA Space Weather Prediction Center) is the WSA-ENLIL model \citep[e.g.,][]{odstrcil05}. ICMEs (with no magnetic field of their own) are simulated in this model by specifying a cone of constant velocity at the inner boundary of a heliospheric domain. In the foreseeable future, however, the steadily increasing computational capabilities may allow the use of thermodynamic MHD simulations for real-time space-weather predictions \citep{jin.m17a}. It is thus important to continuously improve the capabilities and accuracy of these simulations. 

Here we describe a coupled thermodynamic-heliospheric MHD simulation of the 14 July 2000 Bastille Day solar eruption. The simulation covers the evolution of the event from its pre-eruptive state low in the corona to the arrival of the ICME at 1\,AU. The Bastille Day event was one of the strongest eruptions of solar cycle 23 (see \S\,\ref{s:obs}); it thus provides an excellent case for testing the ability of MHD simulations to reproduce the observed properties of extreme eruptions. 

An important feature of our simulation is the construction of a pre-eruptive configuration that is in stable magnetic equilibrium. This extends other thermodynamic MHD simulations of observed eruptions, which typically insert a magnetic flux rope that is not in magnetic equilibrium into the background corona to initiate a CME \cite[e.g.,][]{manchester08,lugaz11,jin.m17b}. This ``out-of-equilibrium'' approach is technically convenient, as it reduces the complexity of the computation while reproducing many observed CME/ICME properties sufficiently well. It is therefore a reasonable approach for trying to develop operational space-weather forecast via MHD simulations in the near future. However, in the longer term, a physically better constrained modeling of CMEs, starting from a stable magnetic equilibrium, is preferable for several reasons. 

First, large-scale solar eruptions always originate from stable configurations, so any realistic model should aim to reproduce this property. Second, producing such configurations allows one to (1) directly compare the model with observations of the pre-eruptive source region; (2) provide a better understanding of how, and how much, free magnetic energy is stored prior to eruptions (while out-of-equilibrium eruption simulations, which do not use such constraints, may release significantly more magnetic energy than is actually available); (3) apply different physical mechanisms for the triggering of CMEs; (4) model the slow rise phase preceding many eruptions \citep[e.g.,][]{liu.r12a}; (5) model cases in which the eruption proceeds successively along the PIL (e.g., \citealt{liu.r09}; see also \S\S\,\ref{ss:setup_AR}--\ref{ss:simu_onset});  and (6) simulate the early kinematic and dynamic evolution of CMEs, as well as phenomena associated with this early phase (e.g., shock-formation low in the corona, EUV waves, dimmings), in a more realistic manner \citep[e.g.,][]{downs11}. Specifically, out-of-equilibrium flux ropes, due to their immediate expansion, may not well reproduce the frequently observed, $B_z$-relevant rotation of CMEs about their rise direction \citep[e.g.,][]{demoulin08,thompson12}. This is particularly true for cases in which most of the rotation occurs low in the corona, while the ejected flux is still accelerating \citep[e.g.,][]{torok10,kliem12,fan16}. Furthermore, inserting a flux rope into a background corona inevitably triggers an unphysical, wave-like perturbation of the system, which superimposes with the modeled CME if the latter starts immediately. While such a perturbation may be damped to some degree by adding sufficient mass to the rope, it is preferable to let it propagate away from the source region before a CME is initiated. 

Finally, present MHD simulations that aim to reproduce (or predict) the {\em in situ} measurements of an actual ICME rely on observations of the associated eruption. Typically, the observed propagation speed of the CME in the corona is required, which is then used to impose an appropriate velocity perturbation to the system \citep[e.g.,][]{odstrcil05,shen.f14} or to constrain the initial flux-rope parameters \citep{jin.m17a}. In contrast, simulations that start from a stable pre-eruptive configuration produce CME properties such as the field strength, orientation, and speed self-consistently, so that no observations of the actual eruption are required to set up the simulation.\footnote{In the case presented here we used the observed flare arcade to constrain the initial flux-rope configuration and the flows used to trigger the eruption (\S\S\,\ref{ss:setup_AR}--\ref{ss:simu_onset}) However, such simulations can be set up by using only pre-eruptive observations such as the location of filaments, sigmoids, or shear along the PIL obtained from vector data.} This allows one to predict the properties and impact of a potential CME {\em before} an actual eruption has occurred, which, in turn, provides the means to assess the strength and impact of eruptions that a given source region on the Sun may produce.

A concise overview of our simulation was given in \cite{linker16}; here we present a more extended description of the numerical setup, the methodology, and the results. This article is organized as follows. In \S\,\ref{s:obs}, we briefly describe the Bastille Day eruption and the associated ICME. In \S\,\ref{s:num}, we present our numerical setup and methodology, with particular focus on the construction of the pre-eruptive configuration. In \S\,\ref{s:cme}, we describe the eruption, discuss the conditions necessary for storing adequate magnetic energy to power such extreme events, and compare the simulation with the observation. \S\,\ref{s:int} addresses the results of our heliospheric simulation and their comparison with {\em in situ} data. We conclude with a summary and discussion in \S\,\ref{s:dis}.

%%%%%%%%%%%%%%%%%%%%%%%%%%%%%%%%%%%%%%%%%%%%%%%%%%%%%%% 
%%%%%%%%%%%%%%%%%%%%%%%%%%%%%%%%%%%%%%%%%%%%%%%%%%%%%%%
%
\section{The Bastille Day Event}
\label{s:obs}
%
%%%%%%%%%%%%%%%%%%%%%%%%%%%%%%%%%%%%%%%%%%%%%%%%%%%%%%%
%%%%%%%%%%%%%%%%%%%%%%%%%%%%%%%%%%%%%%%%%%%%%%%%%%%%%%%
%
The Bastille Day eruption occurred on 14 July 2000 in active region (AR) NOAA 9077. It was one of the largest events during the solar cycle 23. The eruption has been extensively studied, and many articles have been published, including a special volume of Solar Physics (2001, Vol.\,204, Issue 1--2). At the onset of the eruption, the AR was located at disc center, about 15-20\degree\,north of the equator. The event consisted of a filament eruption, an X5.7 flare starting at 10:03\,UT, and a fast moving halo CME with a propagation speed of up to about 1700\,km\,s$^{-1}$ \citep{andrews01}. The flare was followed by an intense radiation storm that resulted in one of the 16 ground level enhancement (GLE) events of cycle 23 \citep{bieber02}. The shock wave driven by the ICME associated with the eruption reached the WIND spacecraft \citep{lin.r.p95,lepping95} and the {\em Advanced Composition Explorer} \citep[ACE;][]{stone98} in $\approx$\,28 hours, followed by a large MC that arrived $\approx$\,5 hours later, around 19\,UT on July 15, with a field strength of $\approx$\,50\,nT and a speed of $\approx$\,1100\,km\,s$^{-1}$ \citep[e.g.,][]{smith01}. The MC, carrying a strong southward magnetic field component (``negative $B_z$''), produced a very strong geomagnetic storm with a minimum geomagnetic storm index, $Dst$, lower than -300\,nT \citep{lepping01}. The {\em in situ} magnetic-field measurements taken at ACE suggest that the spacecraft passed below the axis of a left-handed flux rope \citep{yurchyshyn01}.  

The Bastille Day eruption was followed $\approx$\,3.5\,h later by a second, weaker event that occurred at the western edge of the AR and produced an M3.7 flare \cite[e.g.,][]{andrews01}. This second eruption occurs self-consistently in our simulation, i.e., without the need to impose boundary-driving or  some other external perturbation to initiate it. This suggests that these two eruptions were ``sympathetic'' events (e.g., \citealt{schrijver11,torok11a}; see \S\,\ref{ss:symp}).

We note that the time period around the Bastille Day event was characterized by strong eruptive activity. For example, a large trans-equatorial filament, apparently connected to the eastern section of NOAA AR 9077, erupted almost simultaneously \citep{wang.j.x06}. Moreover, several ICMEs and shocks associated with eruptions were observed by ACE and WIND in the 5-6 days before the MC associated with the Bastille Day CME arrived at the Earth \citep[e.g.,][]{smith01,wang.c01,richardson10}. None of these additional eruptions are included in our simulation. Yet, since they potentially influenced the evolution, trajectory, and final state at 1 AU of the Bastille Day event in unknown ways, their presence has to be taken into account when evaluating the accuracy of our simulation (see \S\,\ref{s:dis}).

%%%%%%%%%%%%%%%%%%%%%%%%%%%%%%%%%%%%%%%%%%%%%%%%%%%%%%% 
%%%%%%%%%%%%%%%%%%%%%%%%%%%%%%%%%%%%%%%%%%%%%%%%%%%%%%%
%
\section{Numerical setup and methodology}
\label{s:num}
%
%%%%%%%%%%%%%%%%%%%%%%%%%%%%%%%%%%%%%%%%%%%%%%%%%%%%%%%
%%%%%%%%%%%%%%%%%%%%%%%%%%%%%%%%%%%%%%%%%%%%%%%%%%%%%%%
%
The coronal evolution of the Bastille Day event (\S\,\ref{s:cme}) was modeled using the thermodynamic MHD code ``Magnetohydrodynamic Algorithm outside a Sphere'' (MAS), developed and maintained at Predictive Science Inc. (see Appendix\,\ref{s:mas} for details). The interplanetary propagation of the associated ICME to the Earth (\ref{s:int}) was modeled using the recently updated heliospheric capabilities of the MAS code \citep{lionello13a}. In this section we summarize our methodology and the numerical setups for the two simulations, paying particular attention to the description of the steps that were used to produce a magnetically stable pre-eruptive configuration.    

%========================================================
% Figure 1: Surface Magnetic Field
\begin{figure}[t]
\centering
\includegraphics[width=1.\linewidth]{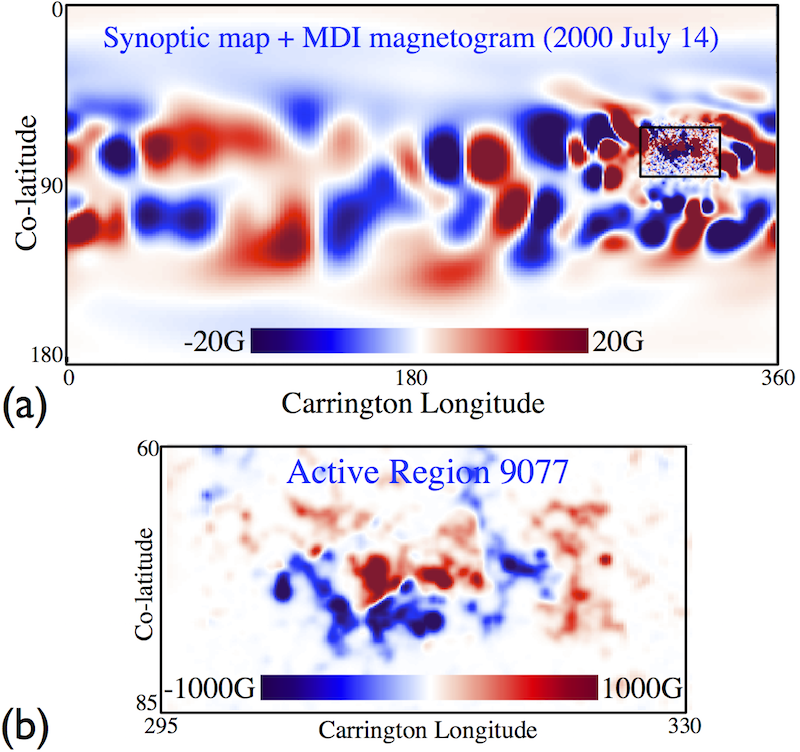}
\caption{
Magnetic map of the radial magnetic field, $B_r$, at the solar surface, used as boundary 
condition in the coronal simulation. 
(a) Full surface field, derived from an MDI synoptic map and a MDI full-disk magnetogram 
(see text for details). 
(b) Close-up view of AR NOAA 9077 (outlined by the small box in (a)), where the Bastille Day 
event and a successive second eruption originated. The maximum magnitude of $B_r$ is 1986\,G.
\label{f:1}
}
\end{figure}
%========================================================

%%%%%%%%%%%%%%%%%%%%%%%%%%%%%%%%%%%%%%%%%%%%%%%%%%%%%%%
\subsection{Global Coronal Background Configuration}
\label{ss:setup_corona}
%%%%%%%%%%%%%%%%%%%%%%%%%%%%%%%%%%%%%%%%%%%%%%%%%%%%%%%
%
In order to provide a realistic background environment for the eruption, we first develop a thermodynamic MHD model of the global corona \citep{lionello09}. To calculate a potential magnetic field that serves as the initial condition for the thermodynamic MHD model, we have to specify the radial field component, $B_r$, at the boundary $r=R_\odot$, where $R_\odot$ is the solar radius. To this end, we combine a line-of-sight (LOS) {\em Solar and Heliospheric Observatory} (SOHO) Michelson Doppler Imager \citep[MDI;][]{scherrer95} synoptic map for Carrington rotation 1965 (July 10\,--\,August 6, 2000) with a LOS MDI magnetogram measured on 14 July 2000 at 09:35\,UT, about half an hour prior to the flare onset (Figure\,\ref{f:1}). To retain as much structure as possible, most of the cells of the numerical grid are concentrated in the source region of the eruption. Outside of the AR, where the grid begins to coarsen, we smooth the synoptic-map magnetic data corresponding to the resolution of the grid. The areas around the poles, where no measurements exist or are not reliable, are fitted in the synoptic map using extrapolation techniques as described in \cite{linker13}.

A particular challenge for the modeling of extreme eruptions such as the Bastille Day event is to construct a pre-eruptive configuration that, on one hand, obeys the observational constraints and, on the other hand, contains a sufficiently large amount of free magnetic energy to reproduce the observed impulsiveness of the eruption and the speed of the associated CME. As demonstrated by \cite{mikic13a}, a prerequisite for achieving this goal is to avoid over-smoothing (or over-diffusing) the observed magnetogram of the source region. Inside the source region, we use a flux-preserving method to resample the LOS MDI magnetogram to a Carrington map with an angular resolution of $\approx$\,0.2 degrees (about the width of 2 MDI pixels). This produces a smooth, but still high-resolution magnetic field when interpolated onto our numerical mesh ($\approx$\,0.1 degrees).

The coronal solution is calculated on a nonuniform spherical ($r,\theta,\phi$) mesh that ranges from 1\,\,$R_\odot$ (the solar surface) to 20\,$R_\odot$ (beyond the sonic and Alfv\'enic critical points). We choose a resolution of $401\times 351\times 471$ mesh points, with $\Delta r$ in the range $4.5\times 10^{-4}R_\odot$ (in the transition region) to 0.4\,$R_\odot$ (at the outer boundary), and latitudinal/longitudinal cells of $\approx$\,0.0017\,--\,0.0019 radians in the area of NOAA AR 9077. The temperature and number density at the lower boundary are fixed to $2\times10^4$\,K and $2\times10^{12}$\,cm$^{-3}$, respectively, similar to the upper chromosphere. The coronal heating function is specified empirically, in a similar manner as in \cite{lionello09} and \cite{downs13}. The parametrization is chosen to give a reasonable match to the observed EUV and soft X-ray emission. The thermodynamic MHD model is calculated for 160 Alfv\'en times, corresponding to about 64 hours (1\,$\tau_A \approx 24$ minutes), until the solar wind has fully opened up the field associated with coronal holes, resulting in a steady-state MHD solution. 

\begin{figure}[t]
\centering
\includegraphics[width=1.\linewidth]{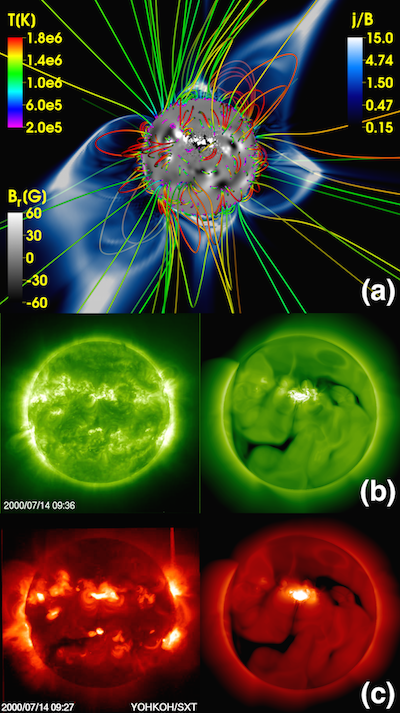}
\caption{
(a) Coronal configuration after the thermodynamic MHD relaxation ($t=160$), seen from Earth 
at $\approx$\,09:30\,UT on 14 July 2000. NOAA AR 9077 is located north of disk center. 
The quantity $|{\bf j}|/|{\bf B}|$ (in code units), where {\bf j} is the current density 
and {\bf B} the magnetic field, is shown in a transparent plane cutting through Sun's 
center; outlining several streamers, i.e., strong activity during the time of the Bastille 
Day event. Field lines are colored by plasma temperature. The radial magnetic field, $B_r$, 
at $R_\odot$ is saturated at 60\,G, to visualize the heavily smoothed ARs outside of NOAA 
AR 9077.    
(b) SOHO/EIT 195\,\AA\,observation of the corona at 09:36\,UT on 14 July 2000, about half 
an hour before eruption (left), and synthetic emission obtained from the simulation at 
$t=162$, after flux-rope insertion and MHD relaxation (right). Emission from smoothed 
and under-resolved ARs outside of NOAA 9077 is not visible in the synthetic image. 
(c) {\em Yohhoh}/SXT observation at a similar time, together with a synthetic {\em Hinode}/XRT 
image at $t=162$.
\label{f:2}
}
\end{figure}

Figure\,\ref{f:2} provides some impressions of the resulting configuration. Panel (a) shows open field lines associated with coronal holes, together with hotter field lines in closed-field areas. Several streamers, visualized by electric currents, can be seen. Panels (b) and (c) show, respectively, the pre-eruptive corona as observed by the Extreme Ultraviolet Imaging Telescope \cite[EIT;][]{delaboudinire95} onboard SOHO and the Soft X-ray Telescope \cite[SXT;][]{tsuneta11} onboard {\em Yohkoh}, together with synthetic emission images obtained from the simulation. One can see that, while observed features such as coronal holes are partly reproduced in the simulation, the overall complexity outside of NOAA AR 9077 is removed. This is because we concentrated the majority of available mesh points in the AR. 

%========================================================
% Figure 2: Global Corona
%========================================================

%========================================================
% Figure 3: Inserted Flux Rope
\begin{figure*}[t]
\centering
\includegraphics[width=1.\linewidth]{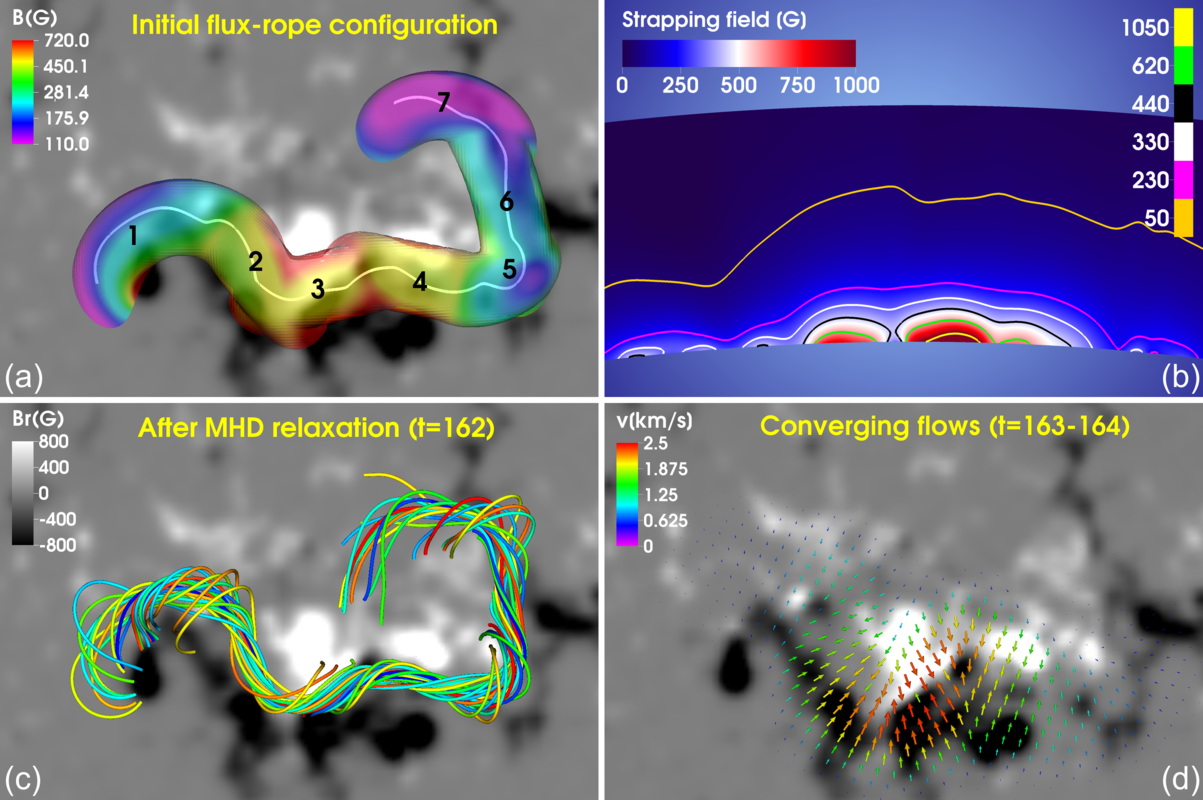}
\caption{
Energization of NOAA AR 9077, the source region of the eruption. 
(a) Flux rope prior to $\beta=0$ relaxation, visualized by a transparent iso-surface of 
$|{\bf j}|=0.04\,|{\bf j}|_{\mathrm{max}}$, colored by magnetic field strength. The white line shows 
the PIL at $r=1.014\,R_\odot$, which runs roughly along the rope axis. Positions of 
individual TDm flux ropes are indicated by numbers. 
(b) Magnetic field component perpendicular to the PIL at $r=1.01\,R_\odot$ (``strapping 
field''), shown in the height range $r=1.0-1.1\,R_\odot$, along a PIL-segment that roughly 
covers TDm ropes 1--5 (from left to right). Colored lines show selected contours.  
(c) Flux-rope field lines after MHD relaxation.
(d) Converging-flow pattern used to trigger the eruption. See text for details.
\label{f:3}
}
\end{figure*}
%========================================================

%%%%%%%%%%%%%%%%%%%%%%%%%%%%%%%%%%%%%%%%%%%%%%%%%%%%%%%
\subsection{Active-Region Energization}
\label{ss:setup_AR}
\begin{figure}[t]
\centering
\includegraphics[width=1.\linewidth]{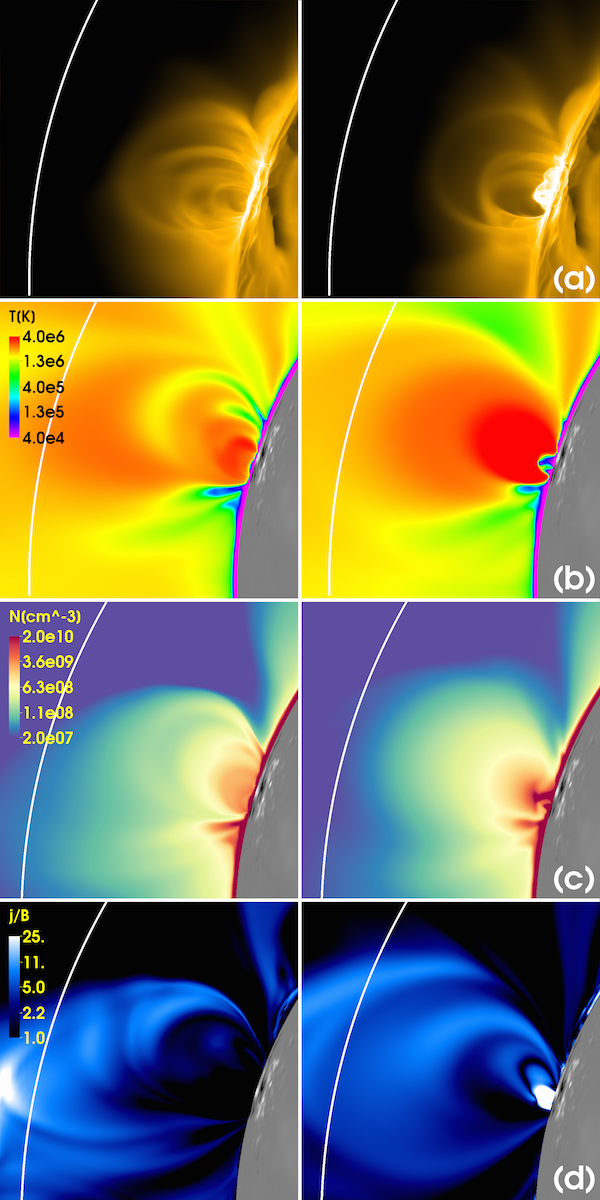}
\caption{
Simulated NOAA AR 9077 before (left; $t=160$) and after (right; $t=162$) 
flux-rope insertion and MHD relaxation, viewed from the West along the main 
(E-W) section of the PIL. The quantities in (b)--(d) are shown in a vertical 
plane roughly perpendicular to the PIL, located approximately in the middle
of the flux rope shown in Figure\,\ref{f:3} (between Tdm ropes 3 and 4).
(a) Synthetic SDO/AIA 171\AA\ emission images.
(b) Plasma temperature.
(c) Number density.
(d) $|{\bf j}|/|{\bf B}|$ (in code units).
The white circle-segment indicates the height $r=1.5\,R_\odot$ (from Sun center)
in all panels.
\label{f:4}
}
\end{figure}

%%%%%%%%%%%%%%%%%%%%%%%%%%%%%%%%%%%%%%%%%%%%%%%%%%%%%%%
%
After the MHD relaxation of the global corona, the magnetic field in the core of NOAA AR 9077 is still relatively close to a potential field (see the left panel in Figure\,\ref{f:4}(d) below). To model an eruption, the AR has to be energized. As discussed in the Introduction, an important aspect of our approach is to start from a pre-eruptive configuration in stable magnetic equilibrium. To this end, we first construct a magnetic flux rope along the eruptive section of the AR's PIL and insert it into the background thermodynamic MHD solution. A simple insertion of a (line-tied) flux rope into a background magnetic field introduces a significant change of the observed magnetogram that was used to calculate the background solution. To avoid such an unphysical perturbation of the system, we use a technique that allows us to insert the flux rope such that the original magnetogram is preserved. Our technique is similar to the one used in the ``Flux-Rope Insertion Method'' \citep{vanballegooijen04}. 

In practice, we separately develop our pre-eruptive flux-rope configuration with the same surface $B_r$ distribution ($B_{r0}$) as the full coronal model (see below). To preserve $B_{r0}$, we calculate the desired flux-rope configuration and compute the photospheric $B_r$ associated with this solution. We then subtract the new $B_r$ distribution from the original $B_{r0}$ and obtain a new potential field. We finally insert the flux rope into this field, which re-introduces the subtracted $B_r$ so that $B_{r0}$ is preserved. This configuration is then numerically relaxed towards a force-free state using a $\beta=0$ MHD model \citep[solution of the momentum equation and Faraday's law with zero plasma pressure, e.g.,][]{mikic94}. Since preservation of the magnetogram changes the potential field into which the flux rope is inserted, several trial-and-error attempts are required until a stable numerical equilibrium is found. The configuration shown in Figure\,\ref{f:3}(c) approached an approximately force-free equilibrium in about 0.2$\,\tau_A$ during the relaxation. 

To insert the relaxed flux rope into the corona, we work directly with the 3D vector potentials, $\mathbf{A}$, that are advanced by MAS (Appendix\,\ref{s:mas}). To get the energized portion of the field only, we calculate $\mathbf{A_{E}}=\mathbf{A_{ZB}}-\mathbf{A_{pot}}$, where $\mathbf{A_{ZB}}$ is the full vector potential of the $\beta=0$ MHD solution, and $\mathbf{A_{pot}}$ is the vector potential of the corresponding potential field. We then add $\mathbf{A_{E}}$ to the full vector potential of the relaxed thermodynamic MHD model of the global corona, $\mathbf{A_{cor}}$. These steps illustrate a useful way to insert energized fields from auxiliary computations or models (in our case the $\beta=0$ flux-rope relaxation) while preserving the surface radial magnetic-field distribution of the global simulation.

To construct our flux-rope configuration, we use the modified Titov-D\'emoulin (TDm) model \citep{titov14}, which is an extension of the original Titov-D\'emoulin (TD) model \citep{titov99}. The latter is an analytical model of a force-free coronal flux-rope equilibrium that has found wide application as an initial condition for CME simulations \citep[e.g.,][]{roussev03,torok05,schrijver08a,kliem10,vandriel14a}. TD flux ropes have been used also in some of the ``out-of-equilibrium'' simulations mentioned in the Introduction \citep[e.g.,][]{lugaz11}. In those cases the stabilizing external ``strapping'' field of the configuration was removed, and the flux rope was inserted into the modeled background field of the eruption's source region. This facilitates the immediate eruption of the rope, but this approach neglects that real eruptions always start from a stable magnetic configuration, in which the Lorentz forces in the pre-eruptive flux are balanced by the background magnetic field.      

In order to provide this capability for CME simulations, the TDm model was designed to facilitate the construction of force-free flux-rope equilibria in an arbitrary background field, as long as that background field is locally (i.e., on the length-scale of the flux rope) bipolar. As in the TD model, the TDm flux rope is a partly submerged torus with constant field strength along its axis. The flux rope is placed in a given background field above the PIL such that its axis approximately follows an iso-contour of the strapping field, i.e., of the component of the background field perpendicular to the PIL (see Figure\,\ref{f:3}(b)). The equilibrium current and axial flux of the rope are then determined using the strength of the strapping field along that contour \citep{titov14}. In order to obtain a stable equilibrium, the thickness (minor radius) and height of the rope apex above the surface must be chosen such that the flux rope is stable with respect to the ideal MHD kink and torus instabilities \citep[e.g.,][]{torok04a,kliem06}. With this technique, flux ropes can be introduced close to, but beneath, the threshold for eruption.  

Before constructing a pre-eruptive configuration, we first have to select the eruptive segment of the PIL of NOAA AR 9077 along which to place a flux rope. For this we use SOHO/EIT and {\em Transition Region And Coronal Explorer} \cite[TRACE;][]{handy99} observations of the flare arcade and filament eruptions (see Figures\,\ref{f:7} and \ref{f:9} below). The two eruptions described in \S\,\ref{s:obs} occurred successively at two adjacent segments of the same PIL (see Figure\,\ref{f:9}(a),(b)). While our main goal is to model the first eruption, we decided to construct a flux rope that continuously covers both segments, rendering the configuration more realistic and allowing us to study the conditions that led to the second eruption. 

Two challenges for constructing a single flux rope along the eruptive segment of the PIL with a single instance of the TDm model are evident from Figure\,\ref{f:3}. First, the segment is very elongated and highly curved (Figure\,\ref{f:3}(a)). Second, the strength of the strapping field strongly varies along it (Figure\,\ref{f:3}(b)). In contrast, the TDm flux rope possesses toroidal geometry (i.e., its axis is straight in projection to the surface) and the field strength is constant along the axial direction of the rope. While the model allows one to construct slightly curved flux ropes with different curvatures via numerical relaxation \citep[see Figures\,3 and 4 in][]{titov14}, it is not flexible enough to be employed for source-region PILs as complex as the one in NOAA AR 9077.\footnote{We have very recently developed a new model, which allows one to construct analytical flux-rope configurations with an arbitrary axis shape \citep{titov18}. This model would have strongly facilitated the construction of the complex pre-eruptive configuration, but it was not yet available when we performed the simulation described here.} 

We therefore use seven individual, overlapping TDm flux ropes, placed as a chain along the eruptive segment of the PIL (Figure\,\ref{f:3}(a)). The chosen size, orientation, inclination, apex height, and field strength of the individual ropes is guided by the strapping-field contours shown in Figure\,\ref{f:3}(b). We calculate the strapping field along the PIL at $r=1.01\,R_\odot$, which roughly corresponds to the height at which we intend to place the axis of the pre-eruptive flux rope. This height is not well constrained by the observations, so the choice of the strapping-field contour along which to approximately place the respective TDm flux-rope axes is essentially a free parameter. Since we aim to maximize the free energy added to the system, we place the ropes relatively low in the corona, where the strapping fields are strong. However, since the ropes need to have a reasonable thickness (10\, Mm or so), we cannot place them too close to the lower boundary. We experimented with different apex heights of the rope axes and found that a height range of $r\approx(1.01-1.015)\,R_\odot$ provides the best compromise. As mentioned above, preserving the original magnetogram required a number of trial-and-error attempts until a stable magnetic equilibrium at this height range was found.  

The axial-field directions of the respective TDm ropes have to have the same sign if they are to merge into a single flux rope. If only line-of-sight magnetograms are available and none of the foot-points of the pre-eruptive flux can be unambiguously associated to a magnetic polarity, the axial-field direction remains a free parameter. Here we choose the direction (top-right to bottom-left in Figure\,\ref{f:3}) such that the flux-rope twist is left-handed (negative helicity), which is suggested by photospheric vector data \citep{zhang.h02}, observed soft X-ray loops and linear force-free field extrapolations \citep{yurchyshyn01}, and observations of the MC associated with the eruption \citep[e.g.,][]{lepping01,yurchyshyn01,lynch05}. The TDm ropes labeled 1--5 in panel (a) cover the area of the main event, while ropes 6 and 7 cover the area of the second eruption. Since all adjacent ropes overlap and their fields superimpose, some axial flux connecting the two endpoints of the structure is present right away, i.e., before any numerical relaxation, while some of the flux has to connect to the surface, due to the strong differences in the field strength of the individual ropes. 

After the insertion of the flux rope into the global background solution we relax the system for about 50 minutes ($t=160-162$). Figure\,\ref{f:3}(c) shows flux-rope field lines after this relaxation (note that the field lines would look rather similar prior to the relaxation in panel (a), where we omitted them for clarity). After the insertion, the resulting configuration is close to being force-balanced but is not in thermal equilibrium. Flows and plasma condensations appear along the flux-rope field lines, possibly arising from thermal non-equilibrium \citep[e.g.,][]{mikic13a}. During the MHD relaxation between $t=160$ and $t=162$, the numerical dissipation (enhanced by the thermal flows) leads to some loss of magnetic energy (see Figure\,\ref{f:5} below). This slow, but continuous loss of magnetic energy was the reason why we did not relax the configuration for a longer time period, which would have allowed the unphysical large-scale wave triggered by the flux-rope insertion to fully leave the numerical domain. At the onset time of the eruption in our simulation ($t\approx164$; see \S\,\ref{s:cme}), this wave has travelled away from the source region to a distance of several $R_\odot$, which is sufficient for avoiding a significant interference of the wave with the actual eruption.   

%========================================================
% Figure 4
%========================================================

Figure\,\ref{f:4} shows the simulated AR before and after the flux-rope insertion and subsequent MHD relaxation, viewed from the West along the main (East-West) section of the PIL. The top panels show synthetic emission images obtained for the simulation for the 171\AA\, passband of the Atmospheric Imaging Assembly \citep[AIA;][]{lemen12} onboard the Solar Dynamics Observatory \citep[SDO;][]{pesnell12}. Several loop-like features are visible in the images \citep[for the formation mechanisms of such loops see, e.g.,][]{mok16}. The remaining panels show the plasma temperature, number density, and the quantity $|{\bf j}|/|{\bf B}|$ (outlining electric currents) in a vertical plane placed at the center of the AR, using the same view that is employed for the emission images. 

Prior to the insertion of the flux rope ($t=160$), the core of the AR is practically current-free and contains hot and moderately dense plasma. Stronger currents are present only at larger heights, around the tip of the streamer that overlies the AR. Surrounding the AR core, collimated regions of relatively cool and dense plasma are visible, resembling the loop-like features in the emission images. The area of very low density north of the AR (dark blue in panel (c)) outlines the base of a coronal hole. After the rope-insertion and subsequent relaxation ($t=162$), the flux-rope current is concentrated in a relatively small area above the surface and the streamer base has expanded, due to the increased amount of closed flux in the AR. The region of hot plasma has expanded as well and its temperature has increased, as a result of additional heating due to the increased magnetic field strength \citep{lionello09}. It can be seen that cold and dense material accumulates just above the PIL, presumably due to plasma evaporation and subsequent condensation \citep[e.g.,][]{xia14}.\footnote{A detailed investigation of the mechanism(s) by which this prominence-like material forms is complex and beyond the scope of this article. It will be the subject of a future publication.} This ``prominence'' material appears bright in the synthetic-emission image, since our optically-thin assumption and procedure of creating such images does not take into account absorption or radiative transfer considerations for material at high densities and low temperatures \citep{mok05}. The plasma beta (the ratio of thermal to magnetic pressure) is small ($\approx$\,$10^{-3}-10^{-2}$) in the AR core, but increases to $\approx$\,1 towards the streamer tip. The Alfv\'en speeds in the AR core are in excess of $10^4$\, km\,s$^{-1}$. Outflows exceeding 100\,km\,s$^{-1}$ are present in the coronal-hole region north of the AR, while the flows are only a few km\,s$^{-1}$ in the AR core.

%%%%%%%%%%%%%%%%%%%%%%%%%%%%%%%%%%%%%%%%%%%%%%%%%%%%%%%
\subsection{Initiation of the Eruption}
\label{ss:simu_onset}
%%%%%%%%%%%%%%%%%%%%%%%%%%%%%%%%%%%%%%%%%%%%%%%%%%%%%%%
%
After the flux-rope insertion and subsequent relaxation, we trigger the eruption of the rope by imposing at $r=R_\odot$ localized, sub-alfv\'enic plasma flows that converge towards the PIL of the source region \citep[e.g.,][]{linker03,mikic13}. Such flows slowly expand the field overlying the flux rope and lead to flux cancellation at the PIL. Both effects result in a slow rise and successive detachment of the flux rope until it becomes unstable and erupts \citep[see, e.g.,][]{aulanier10}. Flux cancellation was observed at several sites of the Bastille Day event's source region, and has therefore been suggested as a trigger mechanism for the eruption \citep[e.g.,][]{zhang.jun01}. We impose the flows during the time interval $t=162-164$ (for about 50 minutes). At $t=164$, when the configuration destabilizes (see \S\,\ref{s:cme}), we ramp down the flows linearly to zero within $0.05\,\tau_A$. 

The imposed flow-pattern is shown in Figure\,\ref{f:3}(d). The pattern is developed ad-hoc, i.e., it is not derived from actual flows measured on the Sun. However, we constrain the flow-pattern by the observed evolution of the Bastille Day event, which started with the eruption of a filament at the western end of the main East-West section of the PIL and then proceeded towards the East, successively building the flare arcade shown in Figures\,\ref{f:7} and \ref{f:9}. 

To this end, we adjust the flows such that the flux rope lifts off first at the section labeled by 5 in Figure\,\ref{f:3}(a), while the sections 6 and 7, which are associated with the second eruption, remain largely unaffected. In our first attempts we repeatedly found that the central part of the rope (sections 4 and 5) indeed erupted, while its eastern part (sections 1--3) did not. 
This happened because, in the narrow strong-field area at the interface of sections 3 and 4, field lines slowly, but continuously disconnected from the rope and became attached to the strong flux concentrations at both sides of the PIL. Then, when the eruption set in, the rope essentially split into two parts, one erupting, the other staying. The slow disconnection of the flux rope started already during the relaxation phase, indicating that the field strengths chosen for sections 3 and 4 were not sufficiently large to balance in this narrow area the downward directed force resulting from the interaction of the flux-rope current and the strapping field. 

Rather than further optimizing our initial flux-rope configuration, we decided to circumvent this undesired behavior by using a ``two-step'' flow-pattern. We first imposed during $t=162-163$ a flow that was localized in the area below sections 3 and 4, to counteract the splitting of the flux in this area. Afterwards, during $t=163-164$, we imposed the full flow pattern shown in Figure\,\ref{f:3}(d), which now led to the eruption of sections 1--3 as well. We eventually fine-tuned the flows further to achieve a smooth lift-off of the rope, progressing successively from section 5 towards section 1.

%%%%%%%%%%%%%%%%%%%%%%%%%%%%%%%%%%%%%%%%%%%%%%%%%%%%%%%
\subsection{Interplanetary Simulation}
\label{ss:simu_helio}
%%%%%%%%%%%%%%%%%%%%%%%%%%%%%%%%%%%%%%%%%%%%%%%%%%%%%%%
%
Finally, in order to model the interplanetary propagation of our Bastille Day CME to Earth, we couple the coronal simulation to the recently updated heliospheric version of MAS \citep{lionello13a,merkin16}. The heliospheric version of MAS solves a simpler set of the MHD equations that neglects radiative losses, thermal conduction, and coronal heating in the energy equation, in either the co-rotating or inertial frame \citep[see][]{lionello13a}.  In the co-rotating frame, the Coriolis and centrifugal force terms are included in the momentum equation.  The heliospheric domain is advectively dominated (flows are supermagnetosonic),  and is therefore less expensive computationally (as compared to the low corona).  A $1484 \times 272\times 368$ nonuniform spherical mesh extending from $r=19$ to 230\,$R_\odot$ is used.

To assure a smooth transition of the CME from the coronal domain into the interplanetary one, we run the coronal simulation until $t=188$ (about 9.6 hours after onset of the eruption), at which time the bulk CME flux-rope has completely left the coronal domain. We then extract for the whole simulation period ($t=160-188$) the variables {\bf B}($t$), {\bf v}($t$), $\rho(t)$, and $T(t)$ at $r=19\,R_\odot$, which are used to drive the interplanetary simulation. 

We start by calculating a potential field in the heliospheric domain based on $B_r(t=160)$. Then, using the variables $B_r$, $v_r$, $\rho$, and $T$ at $t=160$ as fixed boundary conditions in the co-rotating frame, we relax the interplanetary system for 800 $\tau_A$, until a steady state is reached and the Parker spiral has formed, and reset the time to $t=160$. This frame is advantageous as it allows us to concentrate the grid points in the heliospheric domain near the Sun-Earth line during the relaxation. To model the propagation of the CME, we transform to the inertial frame and impose the variables $B_r(t)$, $v_r(t)$, $\rho(t)$, and $T(t)$ as time-dependent boundary conditions (now rotating with the solar rotation rate) at the inner boundary for the whole extracted period, $t=160-188$. The remaining components of ${\bf B}(t)$ and ${\bf v}(t)$ are used for the calculation of the electric fields at the inner boundary, which determines the evolution of the tangential magnetic field. Finally, for $t>188$, only the values of $B_r$, $v_r$, $\rho$, and $T$ extracted at $t=188$ are prescribed as fixed boundary conditions until the end of the simulation, while the remaining components of {\bf B} and {\bf v} are not included in the computation of the electric fields at the inner boundary. For more details on the coupling of the two codes we refer the reader to \cite{lionello13a}.

%%%%%%%%%%%%%%%%%%%%%%%%%%%%%%%%%%%%%%%%%%%%%%%%%%%%%%%
%%%%%%%%%%%%%%%%%%%%%%%%%%%%%%%%%%%%%%%%%%%%%%%%%%%%%%%
%
\section{Results: Coronal Eruption}
\label{s:cme}
%
%%%%%%%%%%%%%%%%%%%%%%%%%%%%%%%%%%%%%%%%%%%%%%%%%%%%%%%
%%%%%%%%%%%%%%%%%%%%%%%%%%%%%%%%%%%%%%%%%%%%%%%%%%%%%%%
%
As a result of the converging flows, the flux rope starts to rise slowly until, shortly after $t\approx164$, it begins to rapidly accelerate upwards, producing a fast CME and a flare. In this section we first discuss the energy evolution of the whole eruption (\S\,\ref{ss:energy}), followed by descriptions of the evolution of the system during the rapid acceleration phase in the low corona (\S\,\ref{ss:early}). We then investigate the propagation of the CME in the outer corona (\S\,\ref{ss:comparison}) and the EUV wave and dimmings associated with the CME (\S\,\ref{ss:wave}), and compare both with available observations. Finally, we describe a second sympathetic eruption that originates in the western section of NOAA AR 9077 (\S\,\ref{ss:symp}).

%%%%%%%%%%%%%%%%%%%%%%%%%%%%%%%%%%%%%%%%%%%%%%%%%%%%%%%
\subsection{Energetics}
\label{ss:energy}
\begin{figure}[t]
\centering
\includegraphics[width=1.\linewidth]{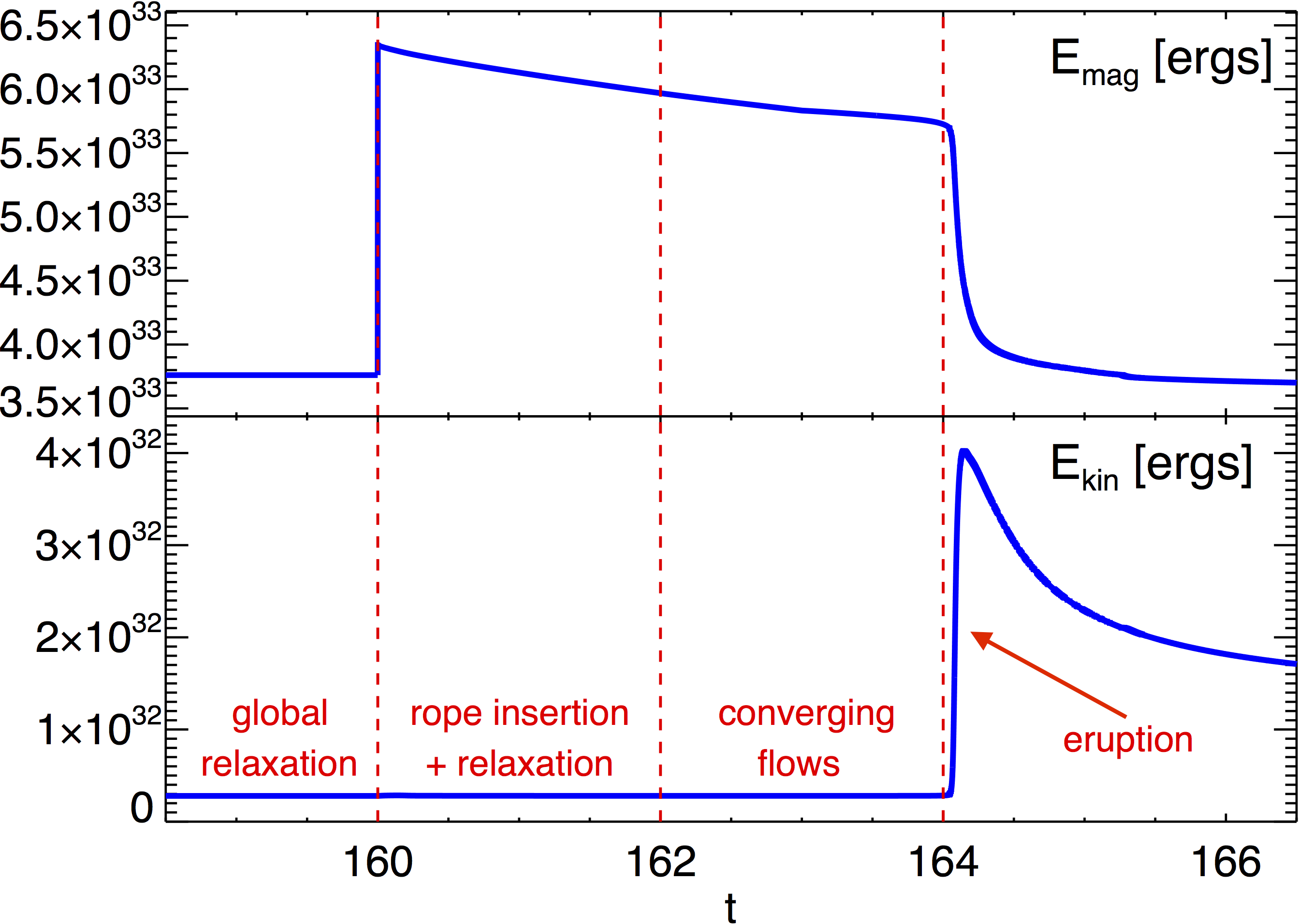}
\caption{
Magnetic and kinetic energies in the simulation around the eruption. 
Times are in $\tau_a$ ($\approx$\,24 minutes). The MHD relaxation 
of the coronal model lasts from $t=0-160$. After insertion of the TDm flux ropes the system is 
relaxed further until $t=162$, then converging flows are imposed during 
$t=162-164$. The system destabilizes shortly after $t=164$.
\label{f:5}
}
\end{figure}

%%%%%%%%%%%%%%%%%%%%%%%%%%%%%%%%%%%%%%%%%%%%%%%%%%%%%%%
%
Figure\,\ref{f:5} shows the evolution of the magnetic and kinetic energies in the coronal domain before and after eruption. The insertion of the flux rope at $t=160$ adds $2.6\times10^{33}$\,ergs of free magnetic energy to the system, a considerable fraction of which is lost due to diffusion and thermal flows during the subsequent relaxation and converging-flow phases between $t=160$ and $t=164$ . After a period of slow rise of the flux rope during the converging-flow phase, the rope rapidly accelerates and the full eruption commences just after $t=164$, releasing 1.3$\times 10^{33}$ ergs in about 4 minutes (0.16$\tau_A$); about 31\% (4$\times 10^{32}$\,ergs) is converted into kinetic energy. This is an important result, as it demonstrates that thermodynamic MHD simulations starting from a pre-eruptive configuration in magnetic equilibrium can reproduce the strong and fast energy release observed in extreme events \citep[this was not clear previously; see the discussion in][]{mikic13}. 

%=========================================================
% Figure 5: Energy evolution
%=========================================================

%========================================================
% Figure 6: Erupting flux rope and flare current sheet
\begin{figure*}[t]
\centering
\includegraphics[width=0.948\linewidth]{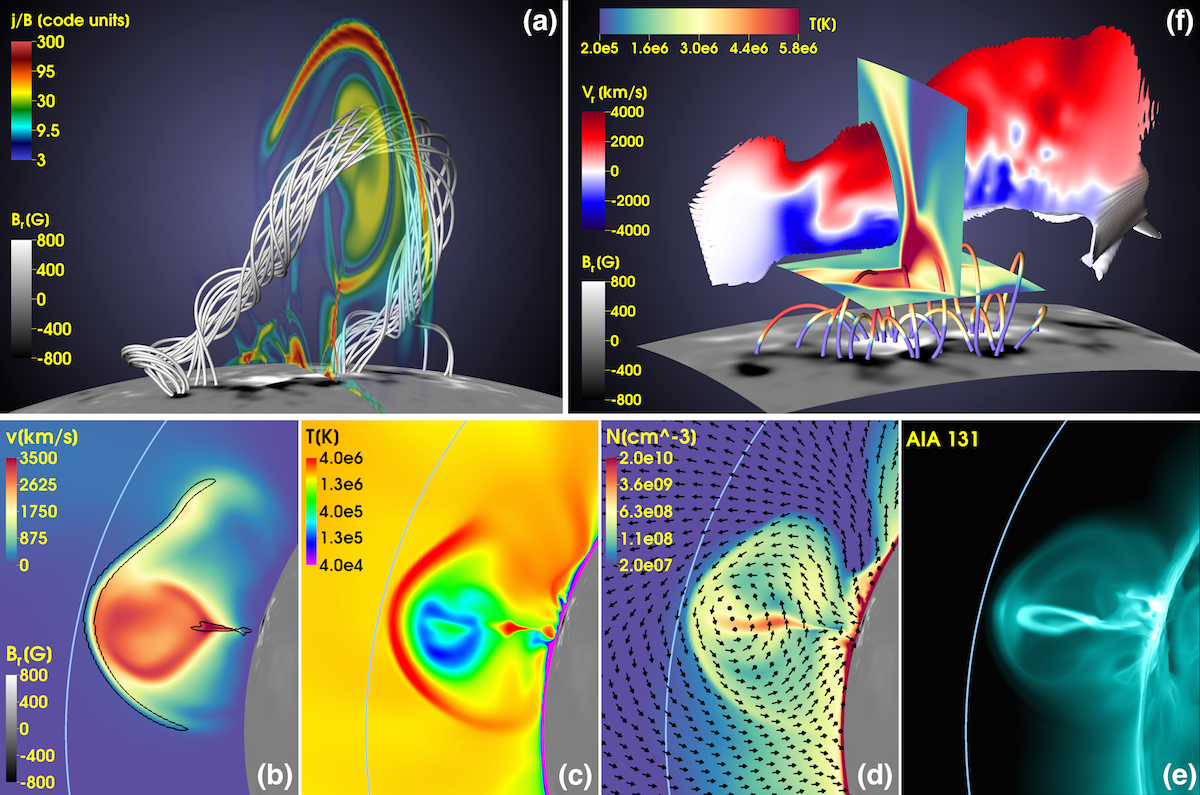}
\caption{
Various quantities during the early eruption phase.
(a) Field lines outlining the flux-rope core at $t=164.10$, shortly after eruption onset. 
The quantity $j/B$ is shown in a transparent vertical plane perpendicular to the rope axis, 
depicting the flux-rope current, the flare-current-layer below the rope, and a compression 
region (shock) in front of the rope.
(b)--(d) Plasma velocity, temperature, and number density in the same plane as in (a), at 
$t=164.12$, about 30\,s later. The black contours in (b) show $\nabla\cdot{\bf v}=0.025$\,s$^{-1}$; 
the arrows in (d) show the orientation of the magnetic field vector.
(e) Synthetic AIA 131\,\AA\,image. The circle-segments in (b)--(e) mark $r=1.5\,R_\odot$.  
(f) Visualization of reconnection below the erupting flux rope, at $t=164.24$, about 3.4\,min 
after the time shown in (a). The main section of the elongated flare-current-layer is shown 
as an iso-volume of $|{\bf j}|/|{\bf B}|$, colored by $v_r$ (depicting reconnection outflows).
Plasma heating and flare ribbons are visualized using two plane-segments that show the plasma
temperature. Reconnected field lines, colored using the same temperature scale, are shown below
the current layer. The center of the layer is removed, to allow the high-temperature cusp in
the vertical plane-segment to be visible. 
\label{f:6}
}
\end{figure*}
%========================================================

It is instructive to compare the energies stored and released in the simulated AR to the AR potential field and open field energies $W_\mathrm{pot}$ and $W_\mathrm{open}$, respectively.  (For a given photospheric flux distribution, the open field is the magnetic field with all field lines starting at the photosphere and extending to infinity.)  Previously, it has been argued \citep{aly91,sturrock91} that the energy of a force-free field cannot exceed $W_\mathrm{open}$. If this conjecture is correct, it places an upper limit on the amount of free energy that can be stored, $W_\mathrm{open}-W_\mathrm{pot}$. We refer to this energy difference as the Maximum Free Energy or MFE. Using the area shown in Figure\,\ref{f:1}(b) to compute $W_\mathrm{pot}=1.71\times10^{33}$\,ergs and $W_\mathrm{open}=5.76\times10^{33}$\,ergs for AR 9077, we obtain MFE $=4.05\times10^{33}$\,ergs. The energy added to the simulated AR when the flux rope is introduced is 2.6$\times10^{33}$\,ergs, or 64\% of the MFE. Some of this energy was reduced during the relaxation and converging-flow phases (the associated flux cancellation also reduces $W_\mathrm{open}$), but 32\% of the original MFE is released impulsively. Our results thus suggest that the MFE may be a good estimate for the maximum energy release of very large events.

%%%%%%%%%%%%%%%%%%%%%%%%%%%%%%%%%%%%%%%%%%%%%%%%%%%%%%%
\subsection{Early Eruption-Phase}
\label{ss:early}
%%%%%%%%%%%%%%%%%%%%%%%%%%%%%%%%%%%%%%%%%%%%%%%%%%%%%%%
%
Figure\,\ref{f:6} shows some impressions of the early phase of the eruption. As discussed above, the rope does not erupt simultaneously, as in out-of-equilibrium simulations. Rather, it successively lifts off from the West to the East within a time period of a few minutes. Figure\,\ref{f:6}(a) depicts the flux-rope core in the course of the rapid acceleration phase, after the eruptive part of the flux has fully disconnected from the surface. The rope does not ascend fully radially, but a few degrees towards the South, due to a slight North-South asymmetry of the ambient magnetic field in the source region. At the time shown in Figure\,\ref{f:6}(a), a part of the western leg of the rope has already reconnected with the ambient field, leading to a displacement of its foot point. This reconnection occurs across a current layer associated with the pseudostreamer located next to the rope (see \S\,\ref{ss:symp} and Figure\,\ref{f:9} below). During this phase, the flux rope accelerates quickly to a speed of $\gtrsim$\,2500\,km\,s$^{-1}$, which leads to a strong compression of the plasma and the magnetic field in front of it and, since its speed exceeds the local Alfv\'en speed, also to the formation of a shock. The shock forms low in the corona, below a height of 1.5\,$R_\odot$. It is visible in Figure\,\ref{f:6}(a) as a layer of strong current density surrounding the flux rope. Such low-coronal shocks (or plasma and magnetic field compressions) are a preferable site for the efficient acceleration of particles to high energies \citep[e.g.,][]{schwadron15a}. It can also be seen that a flare-current-layer has started to form below the rising rope.

%========================================================
% Figure 7: Arcade + Halo CME + height-time
\begin{figure*}[t]
\centering
\includegraphics[width=0.976\linewidth]{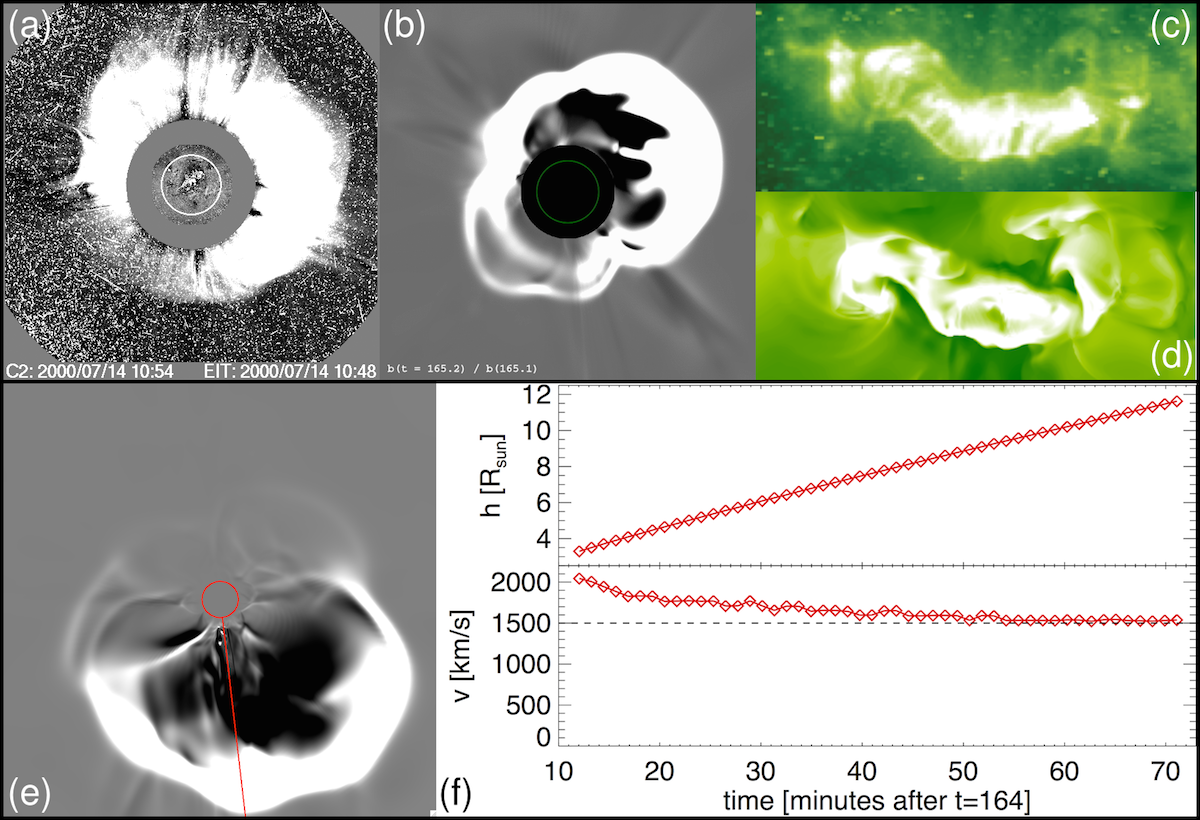}
\caption{ 
(a) SOHO/LASCO C2 difference image of the halo CME associated with the Bastille Day eruption.
(b) Synthetic coronagraph image obtained from the simulation, showing running-ratio brightness
at $t=165.2$, as viewed from Earth. The field-of-view (FOV) is 1.5-6\,$R_\odot$; the green 
circle marks the solar surface.
(c) Flare arcade as seen by SOHO/EIT in 195\AA\,.
(d) Corresponding synthetic emission image obtained from the simulation.
(e) Same as (b) shown looking down on the ecliptic above the north pole of the Sun, at $t=167$, corresponding to the last data point in (f). The red circle outlines the solar surface; the 
FOV is 1-12\,$R_\odot$.
(f) Height and velocity of the simulated CME leading edge between $\approx(3-12)\,R_\odot$.  
The data points were obtained along the red line in (e), using running-ratio brightness 
images.  
\label{f:7}
}
\end{figure*}
%========================================================

Figures\,\ref{f:6}(b)--(e) show the eruption about 30\,s later, when the shock has almost reached $r=1.5\,R_\odot$ and the EUV wave associated with the eruption has started to disconnect from the flux rope in the northern direction, where the Alfv\'en speed is larger (see also \S\,\ref{ss:wave}). We overlaid the velocities in Figure\,\ref{f:6}(b) with a contour of $\nabla\cdot{\bf v}$, in order to visualize the locations of fast-mode shocks \citep{forbes90}. Apart from the shock in front of the flux rope, two termination shocks form below the rope. Those result from the flare-reconnection outflows, which exceed the CME speed during this phase of the evolution (the outflow speeds locally reach $10^4$\,km\,s$^{-1}$). Figures\,\ref{f:6}(c),(d) show that dense and predominantly cold plasma is carried upwards in the bottom part of the flux rope, the center of which is outlined by those arrows that are pointing out of the plane in Figures\,\ref{f:6}(d). Note that the field direction is approximately parallel in front of the rope, i.e., the current layer that precedes the flux ropes forms due to the rapid compression of the magnetic field rather than predominantly due to the perturbation of a null point or null line \citep[as it is the case in quadrupolar configurations; e.g.,][]{antiochos99a}. The plasma surrounding the flux rope is heated to temperatures of up to $\approx\,$6.3\,MK at the time shown, presumably by compressional heating (joule heating was not included in the simulation; see Appendix\,\ref{s:mas}). The temperature further increases as the eruption evolves, reaching $\approx$\,10.9\,MK (below a height of $r \approx 2\,R_\odot$), after which it starts to decrease. Strong heating occurs also above and below the flare-reconnection region, where the plasma reaches peak temperatures of $\approx$\,9.5\,MK. 

Figure\,\ref{f:6}(e) shows a synthetic AIA 131 \AA\, image for comparison. A bubble-like structure, with its edge corresponding to the density and temperature enhancement in front and at the flanks of the rope, is clearly visible. Note that the bright, loop-like feature in the center is not a so-called ``hot flux-rope core'', which has been frequently observed by AIA in this wavelength \citep[e.g.,][]{cheng.x11b}, since the temperatures in this area are much too cold to produce emission in the hot peak of the response function in this wavelength ($\approx10$\,MK). The feature rather outlines dense and predominantly cold plasma, which has accumulated during the pre-eruptive relaxation phase (cf. Figure\,\ref{f:4}), and is now carried upward by the erupting flux rope. Some of this ``prominence'' material probably originates in the layer of dense and cold plasma that is present close to the surface in our model, lifted upwards by initially low-lying flux-rope field lines. This plasma appears bright in the emission image (see \S\,\ref{ss:setup_AR}). In real observations the feature would appear mostly dark in the image, resembling an erupting filament.     

Figure\,\ref{f:6}(f) shows the configuration a few minutes later. The flux rope has left the field of view at this time. Below the rope, a long and elongated flare-current layer has formed, the shape of which follows the eruptive part of the PIL. The coloring of the current layer depicts the flare-reconnection outflows, which are still several 1000 km\,s$^{-1}$ at this time. Hot reconnected field lines have formed below the reconnection region. They make up the flare arcade shown in Figures\,\ref{f:7}(c) and \ref{f:9}(a) below. A cusp-shaped region of hot plasma is visible in the vertical plane segment, while the horizontal plane segment outlines the location of the flare ribbons.   

%========================================================
% Figure 8: EUV wave & dimmings 
\begin{figure*}[t]
\centering
\includegraphics[width=1.\linewidth]{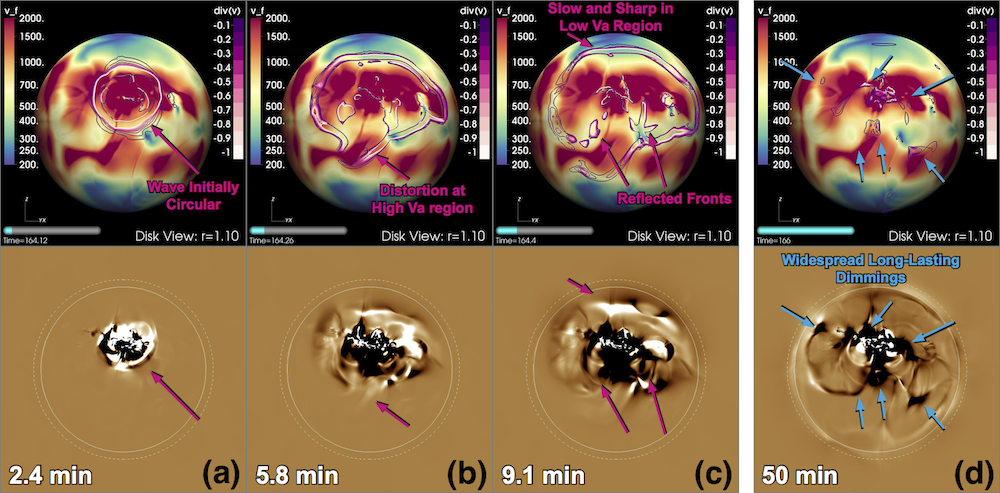}
\caption{
Visualization of the global EUV wave and coronal dimming features present
in the simulation. The top panels show $v_f$ on a sphere taken at 1.1 $R_\odot$, 
with accompanying contours of negative $\nabla\cdot{\bf v}$ (in code units)
to highlight the wave. The bottom panels show synthetic base-difference AIA 
193\AA\ images. The solid and dashed lines in those images indicate distances
of 1.01 and 1.10 $R_\odot$, respectively. 
(a)--(c) Simulation 2.4, 5.8, and 9.1 minutes after the eruption onset; arrows 
indicate interesting wave features, they are at the same positions in the top 
and bottom rows.
(d) Persistent dimming features 50 minutes after eruption onset.
\label{f:8}
} 
\end{figure*}
%========================================================

%%%%%%%%%%%%%%%%%%%%%%%%%%%%%%%%%%%%%%%%%%%%%%%%%%%%%%%
\subsection{Later Eruption-Phase}
\label{ss:comparison}
%%%%%%%%%%%%%%%%%%%%%%%%%%%%%%%%%%%%%%%%%%%%%%%%%%%%%%%
%
In Figure\,\ref{f:7} we compare the simulated eruption with white-light and EUV observations during a later state of the evolution, after the CME leading edge has reached a height of several $R_\odot$, beyond the acceleration region of the solar wind. Figure\,\ref{f:7}(a) shows the large halo CME observed with the Large Angle and Spectrometric Coronagraph \citep[LASCO;][]{brueckner95} onboard SOHO. For comparison, we present in Figure\,\ref{f:7}(b) a synthetic white-light image, obtained by using running-ratio polarized brightness images. A halo-like structure similar to the observed one forms from the simulated CME; the fainter feature seen in the South-East is produced by the large-scale wave associated with the eruption (see \S\,\ref{ss:wave}). Figure\,\ref{f:7}(c) shows emission from the flare in the SOHO/EIT 195\AA\, filter at 11:12\,UT, $\approx50$\,minutes after the maximum flare phase seen in GOES \citep[e.g.,][]{andrews01}. The flare loops were noted for their striking morphology and pattern of growth from West to East, as was also seen in TRACE images. Figure\,\ref{f:7}(d) shows synthetic EIT 195\AA{} emission from the simulation, which reproduces the overall morphology of the flare arcade quite well. The West-to-East growth of the arcade (not shown here) is also reproduced qualitatively. 

Figure\,\ref{f:7}(e) shows the CME in a view onto the ecliptic, at $t=167$, with the direction to Earth being towards the South in the image. We used a sequence of running-ratio brightness images in this view to produce the height-time profile shown in Figure\,\ref{f:7}(f), where the position of the leading edge was always measured along the red line. It can be seen that the CME slowly decelerates from its peak speed of $\gtrsim2500$\,km\,s$^{-1}$ (\S\,\ref{ss:early}) as it travels in the solar wind, until it reaches an almost constant propagation speed of about $1500$\,km\,s$^{-1}$, somewhat less than the estimated peak propagation of speed of $\approx1700$\,km\,s$^{-1}$ \citep{andrews01}. Note, however, that the simulated CME does not move with the same speed in all directions; see also Figure\,5 in \cite{andrews01}. In the low corona, this is due to the presence of background regions with different Alfv\'en speed, and presumably also because the underlying flux rope does not erupt simultaneously along the PIL. The resulting distortion of the CME shape amplifies as the ejecta travels in regions of non-uniform solar-wind speed (see \S\,\ref{s:int} and Figure\,\ref{f:10} below).

%%%%%%%%%%%%%%%%%%%%%%%%%%%%%%%%%%%%%%%%%%%%%%%%%%%%%%%
\subsection{Global Coronal Disturbances}
\label{ss:wave}
%%%%%%%%%%%%%%%%%%%%%%%%%%%%%%%%%%%%%%%%%%%%%%%%%%%%%%%
%
The simulation allows us also to examine the response of the global corona due to the CME. Large-scale propagating coronal waves, or EUV waves \citep[e.g.,][]{warmuth15}, are commonly associated with CMEs in the low corona, and we can clearly identify such a feature in the simulation. The top row of Figure\,\ref{f:8} shows the perpendicular fast-mode magnetosonic wave speed, $v_f=\sqrt{v_a^2+c_s^2}$, where $v_a$ is the Alfv\'en speed and $c_s$ is the sound speed, at a height of 1.1\,$R_\odot$ along with negative contours of $\nabla\cdot{\bf v}$, which is useful for capturing the outer front of a compressible wave \citep[e.g.,][]{wang.h09}. Panels (a)--(c) follow the evolution about 2.4, 5.8, and 9.1 minutes after the eruption. The expansion of the CME introduces a strong, initially circular front, which rapidly distorts according to the local fast-mode speed. In the East and West directions, the front expand much more rapidly because of the persistently large $v_f$, while it slows and sharpens to the North where $v_f$ drops rapidly to quiet-sun values (200--300\,km\,s$^{-1}$). In the South-SouthEast direction, a noticeable distortion is present where the wave expands into a high $v_f$ region, which is adjacent a low $v_f$ region to its west. At later times, secondary fronts or reflections at the high-to-low or low-to-high speed interfaces are also visible. Such behavior, distortions and reflections, are to be expected if the wave propagates at or slightly above the local magnetosonic speed. This result is also consistent with previous simulations \citep[e.g.][]{schmidt10,downs12}.

In the bottom panels of Figure \ref{f:8} we show the corresponding EUV evolution using synthetic AIA 193\AA\, base-difference images. Although the signal is difficult to compare one-to-one to the fast-mode speed at 1.1 $R_\odot$, because of significant projection effects and local variations of the ambient coronal temperature, there is a qualitative correspondence to the outer shape of the 193\AA\, front and the distortions in $\nabla\cdot{\bf v}$ shown in the top panels. Note that in this case, the outer EUV wave front appears dark in AIA 193\AA. This is because the ambient $1.7\!-\!2.0$ MK temperature of the model corona surrounding the erupting AR is slightly above the $\sim\!1.5$ MK peak temperature response of the 193\AA\, channel. This means that the compression and temperature enhancements due to the initial wave passage can lower the LOS emissivity \citep[as discussed in the appendix of][]{downs12}. Such behavior is quite common for EUV waves in the AIA 171\AA\, channel \citep{nitta13}, but is sometimes also observed in the AIA 193\AA\, channel.

For the actual Bastille Day event it was difficult to identify a clear coronal EUV wave, partially due to the 12 minute cadence limitation of EIT and the considerable amount of ``snow'' in the images caused by the strong release of energetic particles during the event. \citet{andrews01} did not find any clear wave signatures but argued that the observed dimmings indicate the presence of a wave, while \citet{chertok05} claimed that a weak wave with a propagation speed of $\approx200$\,km\,s$^{-1}$ was visible to the North-West, where no ARs were present. They mention that projection effects made it difficult to measure the kinematics accurately, but their estimation is roughly consistent with the magnetosonic speed at 1.1\,$R_\odot$ in the quiet-sun region to the North in our simulation. Using height-time base-difference analysis for the northern part of the front in the synthetic EUV images, we estimate a somewhat higher speed of $\approx300$\,km\,s$^{-1}$. 

Lastly, a major discussion of \citet{chertok05} was of the coronal dimmings visible in EIT 195\AA\, at large transverse distances from the erupting AR. These dimmings were visible for at least one hour after the eruption, and might indicate long-lasting density depletion related to the CME or coronal reconfiguration. The bottom panel of Figure \ref{f:8}(d) shows a synthetic AIA 193\AA\, base-difference image about 50 minutes after the eruption (AIA 193\AA\, is very similar to EIT 195\AA\,). This is long after the bulk of the CME has left the low corona, and yet we see long-lasting dimming features in the simulation as well. The extended distribution of these dimmings over the solar disk bears a qualitative resemblance to those shown in Figure\,1 of \citet{chertok05}. A detailed investigation of the relationship of these dimmings to the large-scale magnetic connectivity of the CME and the background corona is beyond the scope of this article; it will be the subject of a future study.

%========================================================
% Figure 9: Sympathetic eruption
\begin{figure*}[t]
\centering
\includegraphics[width=1.\linewidth]{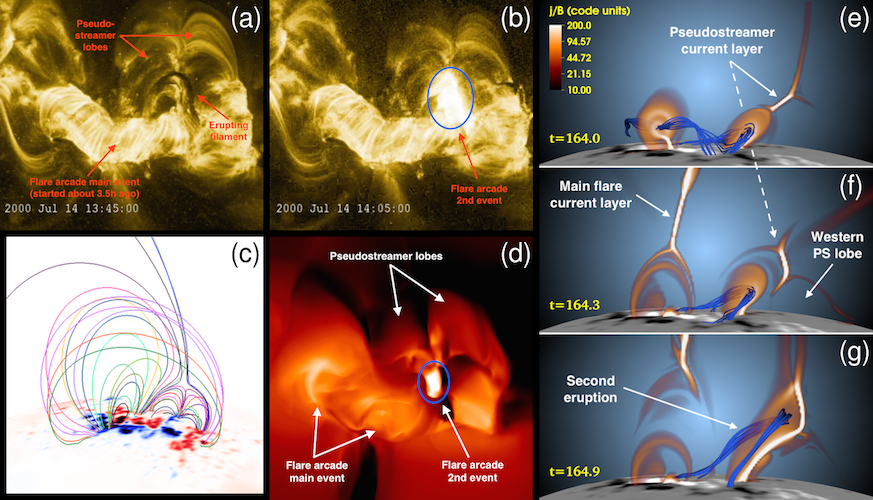}
\caption{
TRACE observations and simulation of the second eruption that occurred in 
NOAA AR 9077 about 3.5 hours after the Bastille Day event.
(a) Filament during its eruption. (b) Flare arcade (encircled) 20 minutes 
later. Loops outlining the PS lobes are visible in both images.
(c): Potential-field extrapolation showing the PS magnetic structure. 
(d): Synthetic {\em Hinode}/XRT Ti\_Poly image during the simulated eruption,
at $t=165.3$. The enhanced flare-emission is encircled. The PS lobes are visible 
in the synthetic emission too.
(e)--(g): Second eruption in the simulation, showing flux-rope segments as 
blue field lines and electric currents visualized by $|{\bf j}|/|{\bf B}|$
(orange-white) in a transparent vertical plane that intersects the pre-eruptive 
flux rope at two locations (see text for details).  
\label{f:9}
}
\end{figure*}
%========================================================

%========================================================
% Figure 10: ICME distortion; run ipbon05a dumps 001, 021, and 049 
\begin{figure*}[t]
\centering
\includegraphics[width=0.913\linewidth]{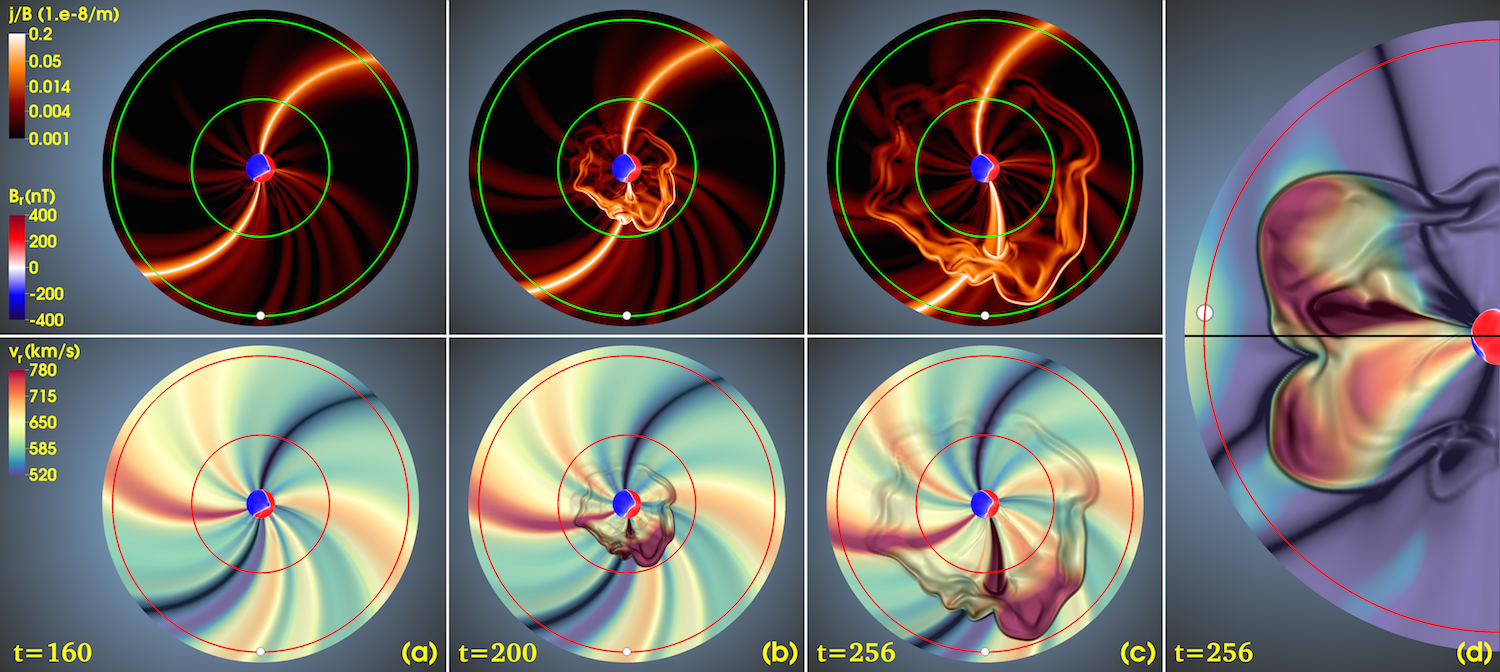}
\caption{
(a)--(c): Electric currents visualized by the quantity $|{\bf j}|/|{\bf B}|$
(top) and radial plasma flows with transparent $|{\bf j}|/|{\bf B}|$ overlaid 
(bottom) in the interplanetary run, shown in the equatorial plane ($\theta=\pi/2$). 
The view is onto the north pole of the Sun. The sphere in the center shows $B_r$ 
at $r=21\,R_\odot$; green and red circles indicate distances of 100 and 
215\,$R_\odot$ from the Sun's center. The little white sphere marks the position 
of the Earth upon the arrival of the ICME core (the magnetic cloud) at 
$r=215\,R_\odot$. The Earth is located 4.5\degree (the $\mathrm{B}_0$ angle 
at this time) above the equatorial plane (see panel (d)). 
(a) Configuration 1.6\,$h$ before the onset of the Bastille Day eruption.
(b) 14.5\,$h$ after eruption onset; the ICME tip has reached $r=100\,R_\odot$. 
(c) 37.0\,$h$ after eruption onset, about 1.2\,$h$ before the ICME tip arrives 
at $r=215\,R_\odot$.
(d) Side view on plasma flows with overlaid $|{\bf j}|/|{\bf B}|$ in the 
$\phi=5.45$ plane, which cuts through Earth's position, 37.0 hours after 
eruption onset. 
%\q{[Maybe submit movie.]}
\label{f:10}
}
\end{figure*}
%========================================================

%%%%%%%%%%%%%%%%%%%%%%%%%%%%%%%%%%%%%%%%%%%%%%%%%%%%%%%
\subsection{Second (Sympathetic) Eruption}
\label{ss:symp}
%%%%%%%%%%%%%%%%%%%%%%%%%%%%%%%%%%%%%%%%%%%%%%%%%%%%%%%
%
The main Bastille Day event was followed about 3.5 hours later by an eruption that took place at the western edge of NOAA AR 9077, along the North-South section of the PIL (roughly at the location of TDm rope 6 in Figure\,\ref{f:3}(a)). It involved a filament eruption and produced an M3.7 flare with onset and peak times at 13:44 and 13:52\,UT, respectively \cite[][]{andrews01}. It is not clear whether a CME was associated with this eruption, since the energetic particles produced by the main event saturated the detectors of the SOHO/LASCO coronagraph until the next day. Figure\,\ref{f:9}(a) and (b) show, respectively, the rising filament and the flare arcade associated with this second eruption. The flare loops of the main event are still visible at this time. Prior to both eruptions, the rising filament may have been connected to the one that took off during the main eruption; this is difficult to deduce from the observations. It was located in the eastern lobe of a pseudostreamer (PS), the two lobes of which are clearly visible as adjacent loop arcades in the TRACE images. The PS is also present in our simulation and depicted in Figure\,\ref{f:9}(c). We note that the PS stalk, while reaching high into the corona, eventually closes back to the solar surface. The structure is therefore not a classical PS, whose stalk would extend into interplanetary space. This difference is, however, not relevant for the eruption scenario discussed in this section. 

It has been shown that PSs provide a favorable environment for ``sympathetic'' eruptions \citep[e.g.,][]{torok11a,panasenco12,titov12,lynch13}. This is because they contain a magnetic null-line above their two lobes at which, when perturbed for instance by an external eruption, a current sheet can form, across which magnetic reconnection takes place. The reconnection transfers magnetic flux from one lobe to the other, which may destabilize a filament (or flux rope) that resides in the lobe whose flux is decreasing. Subsequently, flare reconnection induced by the eruption of the flux rope removes flux from the other lobe, triggering a second eruption, given that a flux rope is present in that lobe as well \citep[for a more detailed description of this process see][]{torok11a}. Note that this mechanism for sympathetic eruptions is not restricted to PSs. It can occur in any magnetic configuration that contains adjacent, closed flux-systems that are rooted in one common polarity and are overlaid by a null line, such as quadrupolar configurations \citep[e.g.,][]{devore05a,peng07,shen.y12,yang.j12,joshi.n16}.  

The relatively small time difference between the Bastille Day event and the second eruption, together with the fact that the filament associated with the second eruption was located in a PS lobe, suggests that these two events were sympathetic, i.e., that the Bastille Day event triggered the second eruption. This is difficult to establish from the observations, but we can turn to the simulation to look for clues. 

Figures\,\ref{f:9}(e)--(g) show snapshots from the simulation at three consecutive times. Figure\,\ref{f:9}(e) shows the AR configuration in a view from the South-West just before the main eruptions starts. The blue field lines show the core of the pre-eruptive flux rope, the eastern part of which is not visible in the computational sub-domain chosen for this illustration (cf. Figure\,\ref{f:3}(c)). The part of the rope located along the East-West section of the PIL is rising upwards, producing the main eruption shortly after. The western part, which is located below the eastern PS lobe, remains at low heights since it is not affected by the converging flows shown in Figure\,\ref{f:3}(d). 

In order to visualize the subsequent evolution, we show the electric currents (visualized by $|{\bf j}|/|{\bf B}|$) in a transparent vertical plane that cuts through the flux rope twice, approximately at the locations where, respectively, the TDm rope-sections 3 and 4 and the TDm rope-sections 6 and 7 intersect (see Figure\,\ref{f:3}(a)). Note that at the time shown in Figure\,\ref{f:9}(e) the orientation of the current layer above the PS lobes is such that reconnection across it would transfer flux from the western PS lobe to the eastern one, i.e., suppressing the second eruption rather than triggering it.

However, as the main eruption proceeds, it initially pushes the PS lobes to the West and downwards. Then, after the erupting flux rope has reached a certain height, the PS lobes expand again and relax back to their old position. During this expansion the orientation of the PS-current-layer reverses, as depicted in Figure\,\ref{f:9}(f). It is this change of orientation that allows the second eruption to occur, since the current layer is now oriented such that reconnection across it, driven by the expansion of the lobes, removes stabilizing flux from the eastern PS lobe. Indeed, as illustrated in Figure\,\ref{f:9}(g), this finally leads to the eruption of the remaining flux-rope section. 

It is important to note that the second eruption occurs self-consistently in the simulation, i.e., without boundary-driving or some external perturbation. The converging flows used to trigger the main eruption are fully switched off at $t=164.05$, i.e., long before the second eruption starts around $t=164.9$. The trigger mechanism of the eruption is the same as described in \cite{torok11a}. Thus, even though the second eruption takes place much earlier in the simulation than in reality (less than 0.5\,h\,vs. $\approx3.5$\,h after the main event), this supports the conjecture that the observed second eruption was sympathetic, i.e., triggered by the Bastille Day eruption. The large discrepancy in the onset time is not surprising; it depends in a very sensitive manner on the detailed structure and evolution of the current sheet and on how the reconnection proceeds, both of which cannot be modeled in a quantitatively correct manner in this type of MHD simulation. We note, for completeness, that \cite{wang.j.x06} suggested that the main Bastille Day eruption was triggered by the almost simultaneous eruption of a large trans-equatorial filament. Since that eruption is not modeled in our simulation, we cannot test this suggestion.
    
%========================================================
% Figure 11: ICME flux rope
\begin{figure*}[t]
\centering
\includegraphics[width=1.\linewidth]{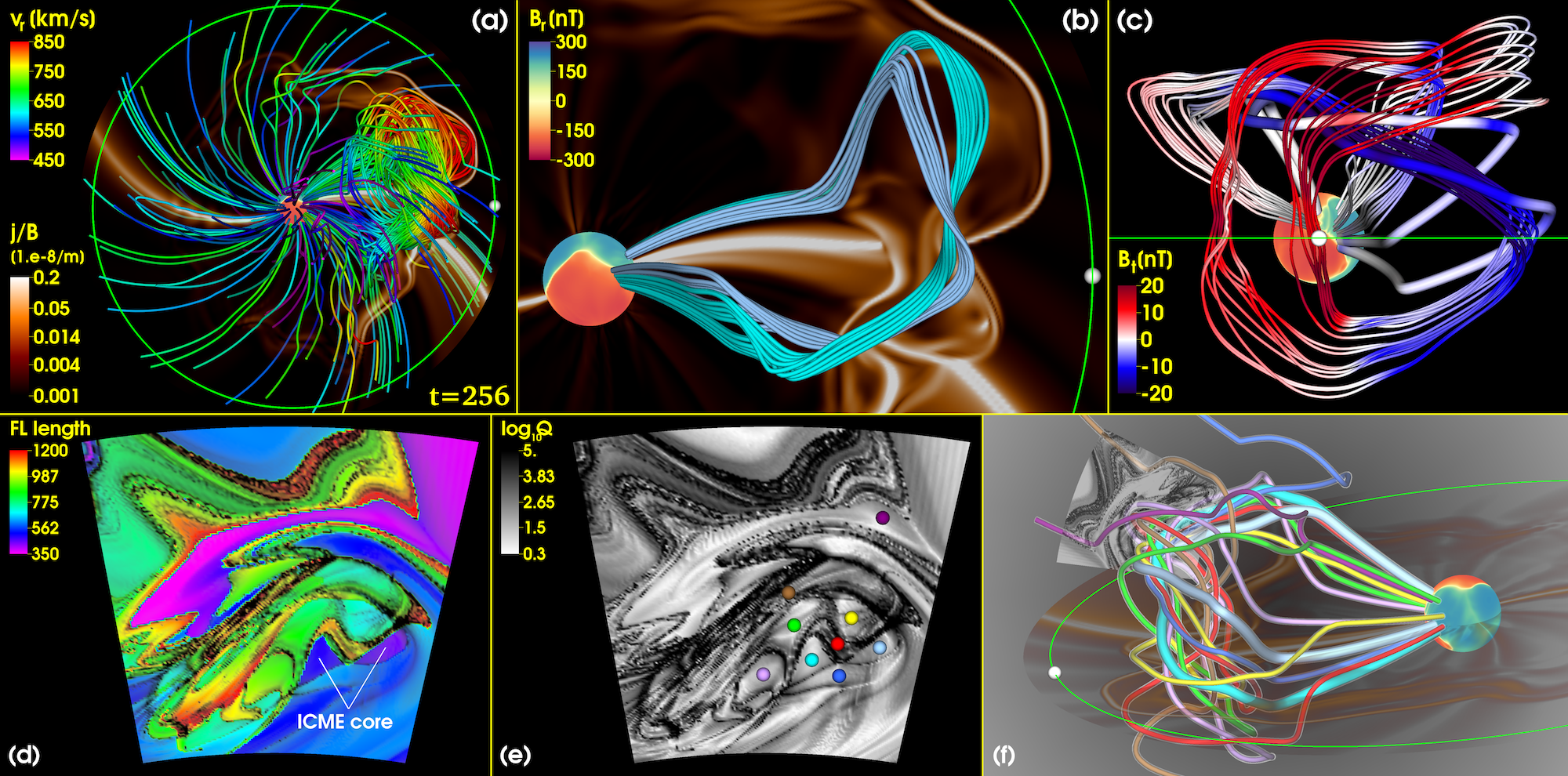}
\caption{
(a) Interplanetary magnetic field and ICME flux rope at $t=256$, shortly before the ICME's arrival at $r=215\,R_\odot$ (green line), with field lines colored by $v_r$. The view is from the North onto the ecliptic plane. The Earth's position (same as in Figure\,\ref{f:10}) is shown by a white sphere. The central sphere shows $B_r$ at $r=20\,R_\odot$. The transparent ecliptic plane depicts electric currents, visualized by $|{\bf j}|/|{\bf B}|$.
(b) Close-up view on (a), showing two flux bundles (cyan and light-blue) that constitute the weakly twisted core of the flux rope.
(c) View from the Earth, with field lines colored by $B_\theta$. Red colors correspond to negative $B_Z$. The flux-rope core is depicted by thick field lines ; thin field lines show two flux bundles that wrap around the core.
(d) Vertical segment extracted at $\phi=5.36$ around the location of the flux-rope core (see (f) for the position of the segment), showing the logarithm of the Squashing factor, $Q$, overlaid on the total field-line length, $L$ (in solar radii).
(e) Same segment, showing only the $Q$-map. Colored spheres mark the positions from which the field lines in (f) were integrated. 
(f) Field lines crossing the $Q$-segment. Thick cyan and light-blue field lines show the ICMe core; thinner field lines cross neighboring areas in the $Q$-map.   
\label{f:11}
}
\end{figure*}
%========================================================

%%%%%%%%%%%%%%%%%%%%%%%%%%%%%%%%%%%%%%%%%%%%%%%%%%%%%%%
%%%%%%%%%%%%%%%%%%%%%%%%%%%%%%%%%%%%%%%%%%%%%%%%%%%%%%%
%
\section{Results: Interplanetary Propagation}
\label{s:int}
%
%%%%%%%%%%%%%%%%%%%%%%%%%%%%%%%%%%%%%%%%%%%%%%%%%%%%%%%
%%%%%%%%%%%%%%%%%%%%%%%%%%%%%%%%%%%%%%%%%%%%%%%%%%%%%%%
%
As described in \S\,\ref{s:obs}, the Bastille Day event was associated with an ICME (and MC) that triggered a large geomagnetic storm at Earth. We modeled the propagation of the ICME in the inner heliosphere as described in \S\,\ref{ss:simu_helio}. In this section we describe the resulting evolution, focusing on the ICME's arrival at Earth.

%%%%%%%%%%%%%%%%%%%%%%%%%%%%%%%%%%%%%%%%%%%%%%%%%%%%%%%
\subsection{ICME Shape and Trajectory}
\label{ss:ICME_shape}
%%%%%%%%%%%%%%%%%%%%%%%%%%%%%%%%%%%%%%%%%%%%%%%%%%%%%%%
%
Figure\,\ref{f:10}(a)--(c) displays the ICME in the equatorial plane at different times of its evolution. Figure\,\ref{f:10}(a) shows the background heliosphere with the Parker spiral, visualized by electric currents (outlining the heliospheric current sheet) and plasma flows. The solar wind is highly structured, i.e., it contains regions of very different plasma-flow speed, with compression regions generated where faster flow follows slower wind, and rarefaction regions where the reverse occurs. The overall solar-wind speed appears to be somewhat large. In our model, it is predominantly determined by the choices made for the coronal heating and the Alfv\'en-wave pressure (see Appendix\,\ref{s:mas}). The former was guided by the observed coronal emission (\S\,\ref{ss:setup_corona}), while the latter was based on our experience from previous solar-wind simulations. Since a direct comparison with the solar-wind conditions at the Earth was not possible due to the strong activity preceding the Bastille Day ICME for almost a week, and since we do not expect that a somewhat slower overall solar-wind would significantly effect the propagation of the ICME, we refrained from experimenting ``blindly'' with these parameters.

It has been suggested that inhomogeneous solar-wind speeds lead to distortions of the ICME shape \citep[e.g.,][]{manchester04b,owens06,savani10}. This indeed happens in our simulation, as inferred from the electric currents in Figure\,\ref{f:10}(b)--(c). Note that the weaker electric currents at the backside of the central sphere (as seen from the Earth) are associated with a large-scale wave triggered by the eruption rather than the actual ICME magnetic field. The most pronounced distortion of the ICME is located to the West of the Earth, corresponding to a part of the ejecta that moves significantly faster than the background solar wind. Adjacent ICME sections do not expand significantly faster than the wind; those show a clear association between shape-distortion and locally enhanced wind speed. Strong deformations of the ICME occur elsewhere in the volume too: for instance, three pronounced ``notches'' can be seen in Figure\,\ref{f:10}(d). Those are at locations where the ICME intersects areas of slow solar wind surrounding the heliospheric current sheet, which is visible in the image as dark ``lanes'' of enhanced current. These deformations of the ICME shape will become important later on, when we discuss the 1\,AU signatures of the ICME and MC derived from the simulation in \S\,\ref{ss:ICME_1AU}. Note in Figure\,\ref{f:10}(c) the presence of an Earth-directed, fast stream behind the ICME front, which is not present prior to the appearance of the ICME. Such streams have been reported to occur frequently behind MCs \citep{fenrich98}. 

The 2D-cuts shown in Figure\,\ref{f:10} do not suggest any significant deflection of the ICME trajectory along the East-West direction \citep[see, e.g.,][]{wang.y02}. According to \cite{wang.y04}, fast ICMEs should be deflected to the East, as a result of pile-up of the magnetic field at their front from the Parker spiral. On the other hand, as those authors pointed out, fast events should be less susceptible to deflection than slow events. The lack of deflection in our case may be due to the relatively fast ICME speed, and possibly also due to the fact that continuous reconnection between the ICME and the background interplanetary magnetic field suppresses a significant flux pile-up.  

We note that deflections of a CME's or ICME's trajectory can occur also due to the interaction with another CME or ICME \citep[e.g.,][]{lugaz12,shen.c12,mishra17}, which is not present in our simulation (see the discussion in \S\,\ref{s:dis}). In such cases, deflection may occur in the North-South direction as well.

%%%%%%%%%%%%%%%%%%%%%%%%%%%%%%%%%%%%%%%%%%%%%%%%%%%%%%%
\subsection{ICME Magnetic Structure}
\label{ss:ICME_field}
%%%%%%%%%%%%%%%%%%%%%%%%%%%%%%%%%%%%%%%%%%%%%%%%%%%%%%%
%
Figure\,\ref{f:11} summarizes the magnetic structure of the ICME flux rope, shortly before it reaches 1\,AU. Figure\,\ref{f:11}(a) shows field lines of the interplanetary magnetic field and the flux rope colored by the radial plasma flow. The overall shape of the rope is distorted (cf. Figure\,\ref{f:10}), and it can be seen that not all parts move at the same speed. The two ``bulges'' at the front of the rope travel with a speed of $\approx$\,850\,km\,s$^{-1}$, significantly faster than the ambient solar wind. 

In order to gain a basic understanding of the magnetic structure of the ICME, we employ the so-called squashing factor, $Q$ (e.g., \citealt{titov02,titov07,pariat12,liu.r16,tassev17}; see Appendix\,\ref{s:Q}), which allows one to identify distinct flux systems \citep[for a similar, much more detailed analysis, see][]{titov17}. We calculate $Q$ in a slice-segment of constant $\phi=5.362$, around the position where our synthetic {\em in situ} data indicate the position of the MC center (20\degree\,North and 5\degree\,East of the Earth's position upon the arrival of the MC at 1\,AU; see \S\,\ref{ss:in-situ}). This segment is shown in Figure\,\ref{f:11}(e); its location in space can be seen in Figure\,\ref{f:11}(f). Dark lines of high $Q$ outline the boundaries between flux systems with different properties. To aid our analysis, we superimpose the Q-map onto an ``L-map'' that shows the total length of the field lines (Figure\,\ref{f:11}(d)).   

Drawing field lines guided by the $Q$ and $L$ maps reveals the presence of two closed, weakly twisted flux bundles of predominantly axial field (Figure\,\ref{f:11}(b)), which we hereafter refer to as the ``core flux''. The location of the core flux regions in the $Q$ map is indicated in Figure\,\ref{f:11}(d). The remaining structure of the flux rope is more complex. In Figure\,\ref{f:11}(f) we show a number of representative field lines that were integrated starting from different regions in the $Q$-map that surround the core flux (their start points are indicated by little spheres in Figure\,\ref{f:11}(e)). Note that the two core-flux bundles are separated from one another by two other flux bundles of different type. Those consist, respectively, of relatively long, closed field lines (red) that wrap around the core flux, and of field lines that apparently were part of the core flux but reconnected with the interplanetary field and are now open (blue). The green, yellow, and light purple field lines are similar to the red one. They all represent closed flux consisting of field lines of rather complex, distorted shape that wrap around the core flux in different ways (note that the red and light purple field lines cross the selecetd $Q$ segment twice). The dark purple field line represents open flux that has been distorted by the ICME (see also Figure\,\ref{f:11}(a)). Finally, the brown field line partially wraps around the core flux, but is open at both ends, i.e., it represents fully disconnected flux that penetrates the ICME flux rope. 
  
Our analysis shows that the basic magnetic structure of the initial coronal flux rope essentially survives during its propagation in the corona and the interplanetary space. However, due to reconnection, likely ongoing at several sites as a result of interaction with the ambient solar wind, the structure becomes increasingly complex. Field lines get distorted into complicated shapes, the core axial flux is intruded by both open and twisted closed flux and eventually splits into two distinct domains, and the flux rope gets pervaded by disconnected flux. A complete understanding of the evolution would be important for improving our understanding of how the flux distribution changes during the propagation of CMEs from the corona to the Earth, but it would require a much more detailed analysis, which is beyond the scope of this article. We leave this to a later study.       
  
Figure\,\ref{f:11}(c) shows field lines colored by the magnitude of $B_\theta$; positive $B_\theta$ (red) is approximately in the direction of negative $B_z$ at the Earth. The two core flux bundles shown in Figure\,\ref{f:11}(b) are depicted here as tubes, and two flux bundles aligned with the azimuthal and wrapped around the core are shown as well. It can be seen that the azimuthal field lines in front of the flux-rope core are dominated by negative $B_z$, i.e., their orientation is favorable for reconnection with the Earth's magnetic field.
  
%========================================================
% Figure 12: 1 AU simu-obs comparison; run ipbon05a 
\begin{figure*}[t]
\centering
\includegraphics[width=0.991\linewidth]{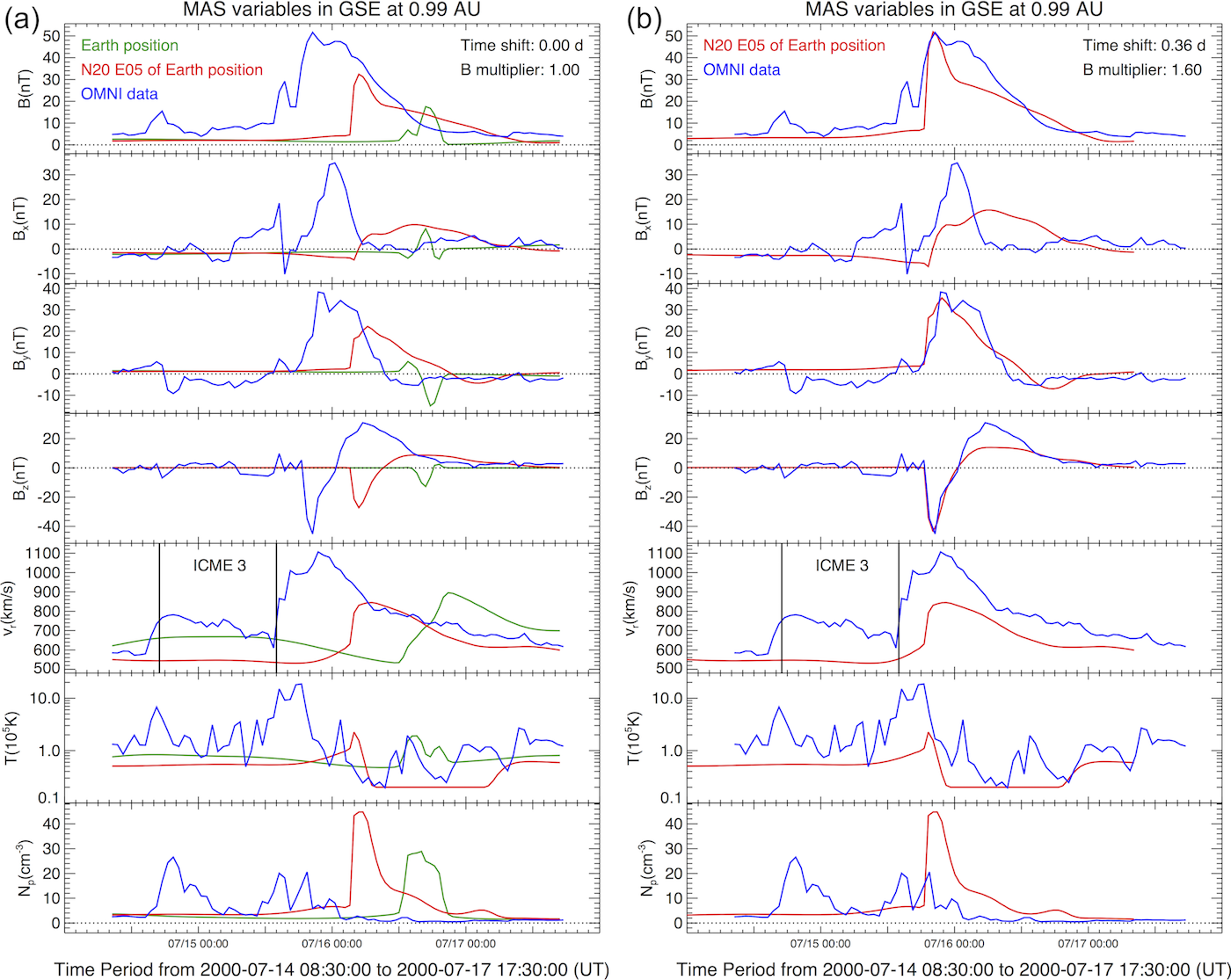}
\caption{
Comparison of real and synthetic {\em in situ} measurements. Blue: one-hour-averaged 
OMNI data in the GSE coordinate system as provided by CDAWeb. Green: simulation data 
at Earth's position at 02:40 UT on 16 July 2000, shortly after the arrival of the
simulated MC at $r=215\,R_\odot$ ($\approx$\,0.99\,AU). Red: simulation data 20\degree\,
North and 5\degree\,East of Earth's position. The vertical lines in the velocity plot
indicate a preceding ICME \citep[termed ICME3 in the notation of][]{smith01}, which is 
not present in our simulation.
(a) Overlay of observations and simulations, without modification of the simulation
data.
(b) Matching the observed and simulated magnetic fields for the MC. To obtain a
reasonable match, the simulation time is shifted by 8.5 hours and the magnetic field
components are multiplied by 1.6.
\label{f:12}
}
\end{figure*}
%========================================================

%%%%%%%%%%%%%%%%%%%%%%%%%%%%%%%%%%%%%%%%%%%%%%%%%%%%%%%
\subsection{Synthetic In Situ Measurements}
\label{ss:in-situ}
\begin{figure*}[t]
\centering
\includegraphics[width=1.\linewidth]{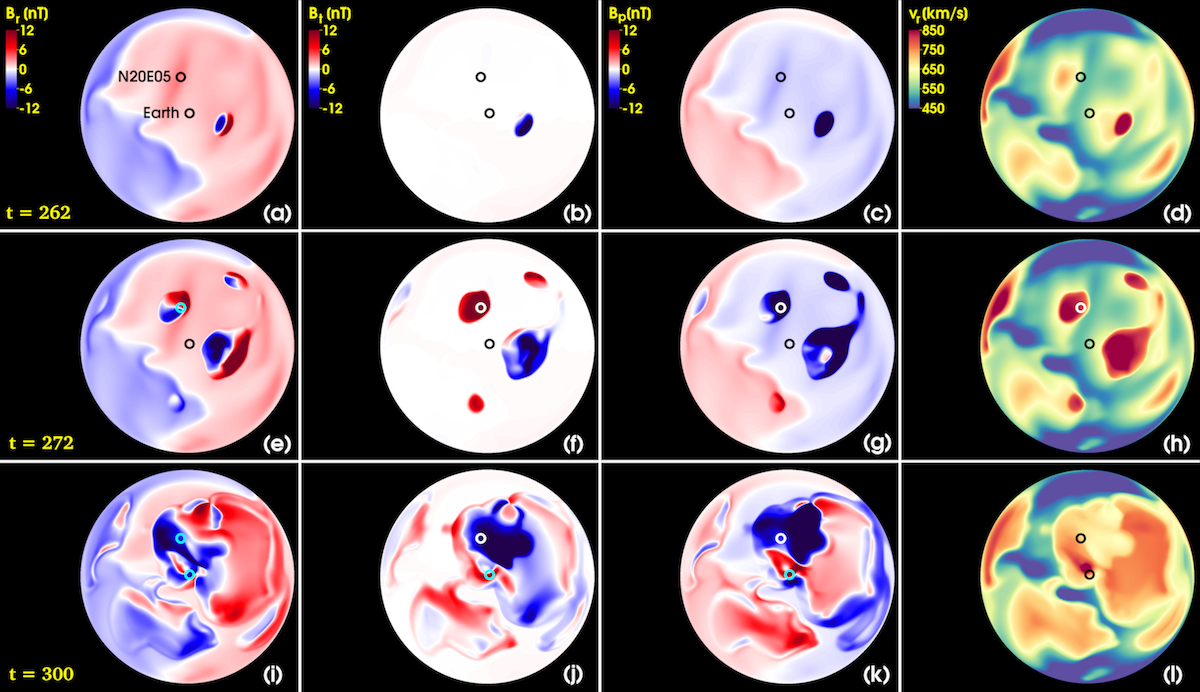}
\caption{
Magnetic field components and radial flow speed ($B_r$, $B_t$, $B_p$, $v_r$;
from left to right) on the surface $r=215\,R_\odot$ at different times in the
interplanetary simulation. The positions at which the green and red curves in
Figure\,\ref{f:12} were obtained are indicated by small circles.
(a)--(d) At $t=262$ ($\approx$\,01:30\,UT on July 16), about one hour after
the first ICME signatures become visible to the West of the Earth.
(e)--(h) At $t=272$, about four hours later, when the magnetic cloud passes
the surface around the location N20E05. 
(i)--(l) At $t=300$ ($\approx$\,16:45\,UT on July 16), when clear ICME
signatures have become visible at the Earth's position. Note the velocity
enhancement just above the Earth's position in panel (l), which corresponds
to the arrival of the high-speed stream behind the ICME (cf. Figure\,\ref{f:10}(c)).
\label{f:13}
}
\end{figure*}

%%%%%%%%%%%%%%%%%%%%%%%%%%%%%%%%%%%%%%%%%%%%%%%%%%%%%%%
%
To compare synthetic {\em in situ} data extracted from the simulation with actual measurements, we first assign a real time to the onset of the simulated eruption in the corona (see \S\,\ref{s:cme}). By comparing the early rise phase of the flux rope shown in Figures\,\ref{f:3}(c) and \ref{f:5}(a) with TRACE observations of the erupting filament, we can associate the simulation time $t=164$ with 10:10\,UT on July 14. The first contact of the simulated ICME with the surface $r=215\,R_\odot$ ($\approx$\,0.99\,AU, where the spacecraft are located) occurs at $t\approx259$, about 38.2 hours after the onset of the eruption (see \S\,\ref{ss:ICME_1AU}). This time corresponds to 00:20\,UT on July 16, which is more than five hours after the arrival time of the observed MC (which was around 19\,UT on July 15; see \S\,\ref{s:obs}). The magnetic field signatures associated with the first contact of the ICME are located $\gtrsim$\,20\degree\,West of Earth, and are dominated by a positive sign of $B_z$ (see Figure\,\ref{f:13}(c) below), i.e., opposite to the sign measured by the WIND and ACE spacecraft at Earth. 

\underline{Synthetic measurements at Earth's position:}
Figure\,\ref{f:12}(a) provides a comparison of synthetic {\em in situ} measurements at the Earth's position (green curves) with one-hour-averaged OMNI data (blue curves). We chose the Earth's position at 02:40\,UT on July 16 for this comparison, which is approximately when the first clear signatures of the MC appear at $r=215\,R_\odot$ in the simulation (red curves; see below for details). The modeled ICME arrives at the Earth's position much later (almost 18 hours) than the observed MC, with a peak speed of roughly 750\,km\,s$^{-1}$, which is about 350\,km\,s$^{-1}$ slower than the measured arrival speed of the MC (the peak velocity of $\approx$\,900\,km\,s$^{-1}$ in the green curve is associated with the high-speed stream that follows the ICME front; see Figure\,\ref{f:10}(c)). The simulated magnetic field strengths at the Earth's position are significantly weaker than the observed ones. While the correct sign of the $B_z$ component is reproduced, the observed rotation of $B_z$ from negative to positive is not. This means that the core of the ICME flux rope (i.e., the MC) does not pass the Earth in the simulation. Figure\,\ref{f:12}(a) also shows that the modeled peak plasma temperature is more than five times smaller than the observed one, while the peak plasma density is about two times larger. This likely indicates that too much cold and dense material is lifted upwards by the low-lying field lines of the flux rope during the coronal eruption (see \S\,\ref{ss:early}).  

\underline{Synthetic measurements at the MC's position:}
Going back to Figure\,\ref{f:11}(c), we see that the core of the ICME flux rope passes the $r=215\,R_\odot$ surface north of the ecliptic (by roughly 20\degree). This was apparently not the case in the real event: while the strong $B_x$ component (see Figure\,\ref{f:12}) suggests that the real MC axis also passed north of the Earth, the pattern of the $B_z$ component suggests that the axis was significantly closer to the ecliptic \citep[see also Figures 9 and 10 in][]{yurchyshyn01}. To infer how the magnetic field components and the plasma quantities would appear in the MC, we add in Figure\,\ref{f:12}(a) synthetic {\em in situ} measurements (red curves) at a location close to the modeled flux-rope axis: 20\degree\,North and 5\degree\,East of the Earth's position (indicated by the circle marked as N20E05 in Figure\,\ref{f:13}(a) below). It can be seen that the synthetic measurements at this location match the observed ones much better than those obtained at the Earth's position: the field strengths are larger and the time difference between the simulated and the real event is much smaller, about 8.5 hours (as mentioned above, the apparently larger speed at the Earth is not associated with the ICME). 

To ease the comparison between the simulated and observed MC signatures, in Figure\,\ref{f:12}(b) we shift the simulation data by 8.5 hours and multiply the magnetic field components by a factor of 1.6, so that the peak magnetic field strengths of the model and the observations match. We can see that the structure of the simulated MC is qualitatively consistent with the cloud inferred from the observations: a full rotation of $B_z$ from negative to positive, and $B_x, B_x > 0$ as the spacecraft passes through the left-handed flux rope below its axis. Specifically, the shape of $B_z$ before the sign switches from negative to positive is almost perfectly reproduced after these modifications, without any stretching of the time axis. We can further see that the shock preceding the observed MC by about 5 hours is not present in the simulation; the shock that forms in the low corona during the early phase of the eruption (see \S\,\ref{ss:early}) has vanished (likely due to numerical diffusion) by the time the ejecta reaches $r=215\,R_\odot$. 

%========================================================
% Figure 13: Surfaces at 0.99 AU; run ipbon05a
%========================================================

%%%%%%%%%%%%%%%%%%%%%%%%%%%%%%%%%%%%%%%%%%%%%%%%%%%%%%%
\subsection{ICME Pattern at 1\,AU}
\label{ss:ICME_1AU}
%%%%%%%%%%%%%%%%%%%%%%%%%%%%%%%%%%%%%%%%%%%%%%%%%%%%%%%
%
Considering Figure\,\ref{f:12}(a) again, it may seem surprising that the green and red curves differ substantially (particularly in the arrival times of the ICME at $r=215\,R_\odot$), given that the transverse distance between their locations is not particularly large: both are well inside the full spatial extent of the ICME (Figure\,\ref{f:13}). The differences are a consequence of the distortion of the ICME shape discussed in \S\,\ref{ss:ICME_shape}. To demonstrate this, we show in Figure\,\ref{f:13} the magnetic field components and the radial plasma flow on the $r=215\,R_\odot$ surface at different times during the passage of the ICME. 

Figure\,\ref{f:13}(a)--(d) shows these quantities at $t=262$ (corresponding to about 01:30\,UT on July 16 in Figure\,\ref{f:12}), about one hour after the first encounter of the ICME with the $r=215\,R_\odot$ surface. A localized bipolar magnetic structure appears approximately 20\degree\,West and 5\degree\,South of the Earth, corresponding to the ICME tip visible in Figure\,\ref{f:10}(c). The structure is dominated by negative $B_\theta$ (panel (b)), i.e., positive $B_z$. Figure\,\ref{f:13}(e)--(h) shows the situation four hours later (about 05:30\,UT on July 16 in Figure\,\ref{f:12}), which is approximately when the simulated MC (which is located around the position N20E05) reveals its peak field strength. Several, clearly isolated signatures of the magnetic field and the plasma flow are visible; the ICME has a surprisingly incoherent appearance. Note, in particular, the very different orientations of the local PILs in panel (e), and the opposite signs of $B_z$ in the MC and the ICME tip, respectively. This scattered and inhomogeneous pattern is associated with the ICME distortion by the solar wind: we see several bulges like those shown in the ecliptic plane in Figure\,\ref{f:10} crossing the surface. Even later, when the main body of the ICME passes through and the distinct patches have merged into a single structure (panels (i)--(l)), the distribution of the quantities over the surface remains rather inhomogeneous. The plasma density and temperature exhibit a similar pattern. We discuss the potential implications of this result in the next section.

%%%%%%%%%%%%%%%%%%%%%%%%%%%%%%%%%%%%%%%%%%%%%%%%%%%%%%%
%%%%%%%%%%%%%%%%%%%%%%%%%%%%%%%%%%%%%%%%%%%%%%%%%%%%%%% 
%
\section{Summary and Discussion}
\label{s:dis}
%
%%%%%%%%%%%%%%%%%%%%%%%%%%%%%%%%%%%%%%%%%%%%%%%%%%%%%%%
%%%%%%%%%%%%%%%%%%%%%%%%%%%%%%%%%%%%%%%%%%%%%%%%%%%%%%%
%
We presented a thermodynamic MHD simulation of an extreme solar eruption: the 14 July 2000 ``Bastille Day'' flare and halo CME. A novel ingredient of our simulation is the initiation of the eruption from a stable magnetic-equilibrium configuration that was constructed using several instances of the modified Titov-D\'emoulin coronal flux-rope model. The simulation reproduces the rapid, strong energy release (about $4\times10^{33}$\,ergs within a few minutes) that is characteristic of extreme eruptions such as the Bastille Day event, which demonstrates, for the first time, that very impulsive eruptions can be modeled with thermodynamic MHD simulations that start from pre-eruptive configurations in magnetic equilibrium in a realistic background. The simulation also  yields good agreement with the observed flare-arcade and halo CME morphologies. The propagation speed of the CME in the outer corona is $\approx$\,1500\,km\,s$^{-1}$, about 200\,km\,s$^{-1}$ slower than the observed peak propagation speed. The simulation self-consistently reproduces a second eruption that occurred in AR NOAA 9077 a few hours after the main event, in one flux lobe of a pseudostreamer-like magnetic configuration located at the western edge of the AR. Our analysis presented in \S\,\ref{ss:symp} suggests that this second eruption was ``sympathetic'', i.e., that it was triggered by the Bastille Day eruption. The physical mechanism that caused the second eruption is similar to the one modeled by \cite{torok11a} and \cite{lynch13}.  

State-of-the-art thermodynamic MHD simulations such as the one presented in this article provide an excellent tool for studying many aspects of solar eruptions in a close-to-realistic magnetic-field and plasma environment. Specifically, they provide information on the magnetic field and the plasma quantities to an extent that is typically not available from observations. Here we investigated, in particular, the EUV wave and dimmings associated with the eruption (\S\,\ref{ss:wave}), but many other topics such as the development of shocks in front of the CME or plasma heating during the flare can be studied with such simulations. 

By coupling our coronal simulation to a heliospheric simulation, we were able to model the propagation of the ICME in interplanetary space to 1\,AU. These kinds of simulations can be employed to study the magnetic structure of ICMEs and MCs, which is helpful for testing and developing 1\,AU flux-rope models. They can also be used to investigate aspects related to the interaction of ICMEs with the solar wind and interplanetary magnetic field, including deflection by the Parker spiral or flux erosion due to reconnection \citep[see, e.g., the recent review by][]{manchester17}.

Here we investigated the magnetic structure of the ICME (\S\,\ref{ss:ICME_field}) and produced synthetic {\em in situ} data that allow a direct comparison to the data measured at 1\,AU (\S\,\ref{ss:in-situ}). We found that the ICME flux-rope core has similar properties as the MC inferred from the {\em in situ} measurements. However, compared to the observed MC, the simulated one has weaker field strengths (by a factor of about 1.6), and it arrives at 1\,AU about 8.5 hours later (with a speed that is about 250\,km\,s$^{-1}$ too low) and about (15-20)\degree\,too far to the north, i.e., it misses the Earth, unlike the observed MC. 

These quantitative differences are not extremely large, given that we used various simplifications, such as a steady-state corona and heliosphere, and did not employ ``trial-and-error'' runs to match the observations (such as the CME speed) quantitatively. Yet, the discrepancies are significant in light of the potential future application of such simulations for space-weather predictions. While the correct sign of $B_z$ at the Earth is reproduced, the observed rotation of $B_z$ is not. Furthermore, the arrival of the ICME at the Earth is delayed by almost 18 hours, and the field strengths are too low by a factor of about three. 

Despite these disagreements, the coronal-heliospheric simulation presented here represent a significant advance in numerical modeling of CMEs and ICMEs, and is among the most detailed and self-consistent simulations of an observed event all the way from its launch at the Sun to its arrival at the Earth. In what follows, we discuss the possible reasons for the discrepancies between the simulation and the real event. 
  
\underline{ICME field strengths:} The simulated field strengths at $r=215\,R_\odot$ are considerably smaller than the measured values. One possible reason for this mismatch could be that the real pre-eruptive core flux was located at lower atmospheric heights (i.e., at locations of larger field strength) than in our simulation. Since we placed our initial flux rope already very close to the bottom boundary, this would imply that the real pre-eruptive core flux must have had a significantly smaller diameter than our rope. Another, more plausible explanation may be an underestimation of the real field strengths in the MDI line-of-sight magnetogram that was used to model the source region of the eruption, NOAA AR9077 (see \S\,\ref{ss:setup_corona}); such underestimates have been described previously by \cite{liu.y07b}. In this respect, it is interesting to note that a wide range of observatory maps appear to underestimate the interplanetary magnetic flux when incorporated into models or coronal hole detections \citep{linker17}, although the reasons for this are uncertain. 

\underline{CME and ICME speed:} Both the simulated CME and MC are about 200 km\,s$^{-1}$ slower than observed ones, which leads to a considerable delay of the ejecta's arrival time at 1\,AU. There may be several reasons for this mismatch. (i) The real ICME was preceded by three consecutive ICMEs that are not included in our simulation. Those may have ``preconditioned'' the solar wind by means of background density depletion, as has been suggested, for example, for the 23 July 2012 event \citep{liu.y14,temmer15}; see also \cite{temmer17a}. (ii) The potential underestimation of the real source-region field strengths may yield an underestimation of the free magnetic energy in the model, and hence an eruption that is not impulsive enough to reproduce the observed CME speed. (iii) The acceleration of CMEs is governed to a large degree by reconnection that takes place below the CME \citep[e.g.,][]{vrsnak08a}. If the reconnection is less efficient in the simulation than in reality, a smaller propagation speed of the CME (and hence of the ICME) may result. (iv) Viscous/resistive dissipation in MHD simulations is greater than in the real solar corona and solar wind, and may lead to artificial slowing of the simulated ejecta.

\underline{CME/ICME-trajectory deviation:} The simulated MC arrives about 15\degree\,too far north compared to the observed one. Again, there may be several reasons for this discrepancy. (i) Our simulation does not include the eruption of a large trans-equatorial filament that took off almost simultaneously with the Bastille Day event \citep{wang.j.x06}. The interaction with this event may have altered the trajectory of the Bastille Day CME \citep[e.g.,][]{lugaz12}. (ii) During the time of the Bastille Day event the flux distribution of the Sun was very complex, with a large number of ARs. Due to computational limitations, the flux outside of NOAA AR 9077 is under-resolved and strongly smoothed in our simulation (see \S\,\ref{ss:setup_corona}). Therefore, large-scale structures such as streamers and coronal holes may not be properly represented in the model. In the real event, such structures may have been deflected the erupting flux rope towards the South. However, it is not clear if that would be a significant effect: strong and fast CMEs like in the Bastille Day event presumably do not get deflected strongly by large-scale coronal structures. Also, a visual inspection of satellite images does not indicate the presence of such structures north of NOAA 9077. (iii) Asymmetries in the magnetic field of a CME source-region can lead to a strongly non-radial rise right at the beginning of the eruption, due to the channeling of the erupting flux by the ambient magnetic field \citep[e.g.,][]{aulanier10,torok13,panasenco13,moestl15,liewer15}. A visual comparison of the simulated early-eruption phase with the TRACE observations indicates that the real CME trajectory was, from the very beginning, directed significantly more southward. As discussed above, strong fields may have been underestimated in the MDI magnetogram we used to model the source region. This may have led to a slight misrepresentation of the magnetic-field asymmetries.           

As described in \S\,\ref{ss:ICME_1AU}, the ICME arrives at 1\,AU with an inhomogeneous pattern of the magnetic field and plasma quantities, with opposite signs of $B_z$ at different locations. We associated this pattern with the distortion of the ICME shape by gradients in the solar-wind speed, though internal reconfigurations of the ICME magnetic field may have played some role as well. 

During solar minima, when the solar wind has a relatively simple bimodal structure, a dominant concave distortion can be expected, at least for low-latitude ICMEs that travel along the heliospheric current sheet \citep[e.g.,][]{odstrcil99,manchester04b,savani10}. During solar maxima, when the structure of the wind is intricate and gradients between regions of fast and slow wind are smaller \citep{riley03b}, ICME distortions will be more complex, but probably less pronounced. However, strong additional (convex) distortions may develop for events in which parts of the ejecta travel significantly faster than the background solar wind (see Figure\,\ref{f:10}(c)).

Therefore, if the pattern shown in Figure\,\ref{f:13} is indeed representative of real (fast) events, this means that a reasonably accurate forecast of observed ICME or MC signatures at the Earth by means of MHD simulations remains very challenging. Inaccuracies in the CME/ICME trajectory of just a few degrees may lead to an incorrect prediction of the sign or magnitude of $B_z$, the speed of the ejecta, and its arrival time. 

Thus, from both a scientific and a space-weather prediction perspective, the MHD modeling of solar eruptions requires further development. As discussed in the Introduction, we believe that initiating CMEs slowly from pre-eruptive configurations in stable magnetic equilibrium is one important step towards more realism and accuracy. The technique presented in \S\,\ref{ss:setup_AR} constitutes a significant step forward, but for complex source regions it may require a significant number of trail-and-error attempts. In order to strongly reduce the number of required trial-and-error attempts, our group has recently developed a generalization of the TDm model, which allows one to use a single flux rope of arbitrary shape for the construction of stable pre-eruptive configurations (Titov et al., in preparation).

The development of such configurations could benefit also from including information from NLFFF extrapolations or models \citep[e.g.,][]{schrijver08b,savcheva09}, or from flux-emergence simulations that model the formation of pre-eruptive configurations \citep[e.g.,][]{archontis08c,cheung10,fang12,leake13,toriumi14}. First steps in this direction have been done by, e.g., \cite{roussev12}, who used a flux-emergence model to drive their CME simulation, albeit only for an idealized magnetic configuration. Also, observed photospheric flows should be included to provide a more realistic description of the energy build-up prior to eruptions and of their initiation \citep[e.g.,][]{jiang.c16}. Furthermore, a more realistic modeling of the environment in which CMEs and ICMEs travel is needed To this end, time-dependent, continuously updated MHD models of the global corona and interplanetary space should be developed, similar to what is already done in NLFFF flux-transport models \cite[see, e.g.,][]{mackay12}. Future models must also be able to simulate several eruptions simultaneously, to account for the interaction of CMEs/ICMEs with one another.

Finally, we would like to note that close-to-realistic simulations like the ones presented here should be employed for detailed investigations of specific aspects of solar eruptions. For example, our thermodynamic MHD simulation data are currently used to evaluate uncertainties in coronal electron temperature and speed measurements (Reginald et al., in preparation), and to model the acceleration and propagation of energetic particles \citep[as described for a different simulation in][]{schwadron14}. We encourage interested researchers to contact us if they would like to use our simulation data for complementary investigations.

\acknowledgments
We thank M. Owens for helpful discussions regarding ICME distortions. This work was supported by AFOSR, the NASA programs LWS C-SWEPA project, LWS team on Flux Ropes, LWS team on interplanetary B$_Z$, and H-SR, and by the NSF programs FESD, SHINE and Solar Terrestrial. Computational resources were provided by the NSF supported Texas Advanced Computing Center (TACC) in Austin and the NASA Advanced Supercomputing Division (NAS) at Ames Research Center.

%%%%%%%%%%%%%%%%%%%%%%%%%%%%%%%%%%%%%%%%%%%%%%%%%%%%%%%
%%%%%%%%%%%%%%%%%%%%%%%%%%%%%%%%%%%%%%%%%%%%%%%%%%%%%%%
%
\appendix
%
%%%%%%%%%%%%%%%%%%%%%%%%%%%%%%%%%%%%%%%%%%%%%%%%%%%%%%%
%%%%%%%%%%%%%%%%%%%%%%%%%%%%%%%%%%%%%%%%%%%%%%%%%%%%%%%

\section{The MAS Thermodynamic MHD Model}
\label{s:mas}
The numerical code MAS employed in this article integrates the standard viscous and resistive one-fluid MHD equations in 3D spherical coordinates. For the coronal simulation described in \S\S\,\ref{ss:setup_corona}--\ref{ss:simu_onset} and \ref{s:cme}, the so-called ``thermodynamic MHD model'' was used, in which the standard equations are extended to include parallel electron thermal conduction, radiative losses, and parameterized coronal heating. The MAS thermodynamic MHD model has been used extensively for simulating the global corona and solar wind \citep[e.g.,][]{mikic99,lionello01,lionello09,mikic07,downs13,titov17,linker17} and eruptive phenomena such as soft X-ray jets \citep{torok16,lionello16} and CMEs \citep[e.g.,][]{linker01,linker03,mikic13}. 
%MAS is also the primary MHD model in CORHEL (Corona-Heliosphere), a suite of models \citep{riley12} for describing the solar corona and inner heliosphere that is available for public use at NASA's CCMC \footnote{\url{http://ccmc.gsfc.nasa.gov/}}. 
In this article, we use a version of the model in which the solar wind is accelerated with Alfv\'en waves using a Wentzel-Kramers-Brillouin (WKB) approximation \citep{jacques77}. A more sophisticated wave-acceleration model is under development \citep{lionello14,downs16}. In the version used here, the governing equations take the following form:
%
%=====================================================================================================
\begin{alignat}{2}
\frac{\partial {\bf A}}{\partial t} &={\bf v}\times \left(\nabla\times {\bf A}\right)-\frac{c^2\,\eta}{4\,\pi}\,\nabla \times \nabla \times {\bf A} \label{eq:mhd:a},
\\
\frac{\partial \rho}{\partial t} &=-\nabla\cdot(\rho\,{\bf v}),
\\
\frac{\partial T}{\partial t} &=- \nabla\cdot(T{\bf v})
-(\gamma - 2)\left(T\,\nabla\cdot{\bf v}\right)-\frac{1}{2}\,\nabla\cdot\left(f_{\mbox{\tiny nc}}(r)\,T\,{\bf v}\,{\bf \hat b}{\bf \hat b}\right) \label{eq:mhd:e1}
\\
&+\left.\frac{(\gamma-1)}{2\,k}\frac{m_p}{\rho}\right[\nabla\cdot\left(\beta_{\mbox{\tiny Tcut}}(T)\,f_{\mbox{\tiny c}}(r)\,\kappa_0\,T^{5/2}\,{\bf \hat b}{\bf \hat b}\cdot\nabla T\right)\left. - \frac{\rho^2}{m_p^2}\,\frac{Q(T)}{\beta_{\mbox{\tiny Tcut}}(T)} + H\right], \notag
\\
\frac{\partial{\bf v}}{\partial t}& = -{\bf v}\cdot \nabla\,{\bf v} + \frac{1}{\rho}\left[ \frac{1}{c}{\bf J} \times {\bf B} - \nabla p - \nabla\left(\frac{\epsilon_{+}+\epsilon_{-}}{2}\right) + \rho\,{\bf g}\right]  
\label{eq:mhd:v}
\\
&+ \frac{1}{\rho}\,\nabla\cdot(\nu \rho \nabla {\bf v})  + \frac{1}{\rho}\,\nabla\cdot\left(S\,\rho\,\nabla{\frac{\partial \bf v}{\partial t}}\right), \notag
\\
\frac{\partial \epsilon_{+}}{\partial t} &=-\nabla\cdot \left( \epsilon_{+}\left[{\bf v}+v_A\,{\bf \hat b}\right]\right) - \frac{1}{2}\,\epsilon_{+}\,\nabla\cdot {\bf v}, \label{eq:mhd:wkb1}
\\
\frac{\partial \epsilon_{-}}{\partial t} &=-\nabla\cdot \left( \epsilon_{-}\left[{\bf v}-v_A\,{\bf \hat b}\right]\right) - \frac{1}{2}\,\epsilon_{-}\,\nabla\cdot {\bf v}, \label{eq:mhd:wkb2}
\end{alignat}
%=====================================================================================================
%
where ${\bf A}$ is the magnetic vector potential, ${\bf B}=\nabla\times {\bf A}$ is the magnetic field, ${\bf J}=\frac{c}{4\,\pi}\nabla\times {\bf B}$ is the current density, $\rho$ is the plasma density, $T$ is the temperature, $p=2\,k\,T\,\rho/m_p$ is the plasma pressure, {\bf v} is the plasma velocity, ${\bf \hat b}=|{\bf B}|/{\bf B}$ is the direction of the magnetic field, $c$ is the speed of light, $\gamma=5/3$ is the adiabatic index, $m_p$ is the proton mass, $k$ is Boltzman's constant, $\kappa_0$ is the coefficient of the classical Spitzer thermal conductivity, $f_{\mbox{\tiny c}}(r)=0.5\,(1-\mbox{tanh}[(r-10R_{\odot})/0.5\,R_{\odot}])$ is a profile that limits the radial extent within which collisional (Spitzer's law) thermal conduction is active, $f_{\mbox{\tiny nc}}(r)=1-f_{\mbox{\tiny c}}(r)$ is the equivalent profile for collision-less thermal conduction, $Q(T)$ is the radiative loss function, $H$ is the coronal heating term that typically consists of a sum of empirical heating functions \citep{lionello09}, $v_A=\sqrt{|{\bf B}|^2/4\,\pi\,\rho}$ is the Alfv\'en-wave speed, and ${\bf g}=-g_0\,R_{\odot}^2\,{\bf \hat r}/r^2$ is the gravitational force. Note that joule heating, $\eta J^2$, is not included in Eq.~(\ref{eq:mhd:e1}). While it is implemented in MAS, we switched it off in our simulation, otherwise the insertion of the current-carrying flux rope into the background corona (\S\,\ref{ss:setup_AR}) would have let to an instantaneous, unphysical temperature increase of the rope. The last expression in Eq.~(\ref{eq:mhd:v}) is a semi-implicit term that is added to the equations to stabilize the algorithm for time-steps larger than the fast magneto-sonic wave limit \citep{lionello99,caplan17}. Eqs.~(\ref{eq:mhd:wkb1}) and (\ref{eq:mhd:wkb2}) are the WKB approximation for Alfv\'en-wave pressure advance \citep{mikic99}, where $\epsilon_{+}$ and $\epsilon_{-}$ are the forward and backward Alfv\'en-wave energy densities. 

The function $\beta_{\mbox{\tiny Tcut}}(T)$ is a cut-off function that serves to broaden the transition region; $\beta_{\mbox{\tiny Tcut}}=(T/T_{\mbox{\tiny cut}})^{5/2}$ for $T<T_{\mbox{\tiny cut}}$ and $\beta_{\mbox{\tiny Tcut}}=1$ for $T \ge T_{\mbox{\tiny cut}}$.  Applying this function allows one to increase the width of the transition region (i.e., its spatial resolution) with a minimal effect on the global coronal solution \citep{lionello09,mikic13a}. Here $T_{\mbox{\tiny cut}}=5\times 10^5 K$ is used. The resistivity, $\eta$, and the kinematic viscosity, $\nu$, are set such that the corresponding diffusion times are $\tau_{\eta}=(4 \pi R_\odot^2)/(\eta c^2) \approx 4\times 10^5$ hours and $\tau_{\nu}=R_\odot^2/\nu \approx 80$ hours, respectively, much larger than the Alfv\'en time of $\approx 24$ minutes. 

For the interplanetary simulation described in \S\S\,\ref{ss:simu_helio} and \ref{s:int}, thermal conduction, radiative losses, and coronal heating are neglected, $\gamma$ is set to 3/2, and a smaller kinematic viscosity, corresponding to $\tau_{\nu} \approx 400$ hours, is used. The characteristic form of the MHD equations is employed for both simulations to specify the boundary conditions at the radial boundaries (see \citealt{linker97} for details).

\section{Analysis of the Magnetic Field Using the Squashing Factor}
\label{s:Q}
Separatrix Surfaces (SSs) and Quasi-Separatrix Layers (QSLs) are, respectively, topological and geometrical features that fully or partly partition magnetic configurations into different flux systems \citep[e.g.,][]{priest14}. They can be identified by computing appropriate maps of the so-called squashing factor (or squashing degree), $Q$ \citep{titov02}. In essence, $Q$ is a measure of how elliptical an infinitesimal circular region of one polarity becomes when mapped along field lines to its conjugate footprint. Its minimum value, $Q=2$, corresponds to the mapped footprint remaining circular. Larger values of $Q$ describe how fanned out a field-line bundle becomes from one end to the other. Regions with $Q >>2$ determine QSLs. In the limit $Q \rightarrow \infty$, occurring at field lines that thread either a magnetic null or a bald patch, the magnetic surface spanned by these field lines becomes a SS. In numerical studies, such a surface appears as unresolved spikes of $Q$, so that both true SSs and QSLs, as well as their hybrids, are detected by computing $Q$ distributions. $Q$ was initially defined for closed field lines, but it can be computed for open field lines as well \citep{titov07,titov08}. Overall, $Q$ becomes very large or infinite at locations where the magnetic structure experiences an abrupt change \citep[e.g.,][]{titov12,titov17,savcheva12}. 

%$Q$ has been shown to have better properties than $N$ for locating QSLs.

By construction, $Q$ is invariant to the direction of the field-line mapping \citep{titov02}. Therefore, its value at conjugate foot points of a field line at a boundary can be assigned to any point of the field line. Thus, by mapping $Q$ along field lines from a boundary (e.g., the inner boundary of our simulation domain) to cross-sections of interest, it becomes possible to visualize QSLs at any plane in the volume \citep{titov08,pariat12,liu.r16}. Such visualization significantly helps to interpret the structure of complex magnetic fields, and it was used in our analysis of the ICME flux rope shown in Figure\,\ref{f:11}. For complex $Q$ maps such as the one shown in Figure\,\ref{f:11}(e), it is useful to color the map by the field-line length, $L$, which yields an ``$L$ map'', as shown in Figure\,\ref{f:11}(d). The $L$ map aids the interpretation, since distinct segments of the same color typically outline flux bundles that cross the $Q$ map two times or more.

\bibliographystyle{yahapj}
\bibliography{/Users/tibor/Desktop/pap/torok}

\listofchanges
\end{document}